\documentclass[paper, nofootinbib]{revtex4}
\usepackage{amsfonts}
\usepackage{graphicx}
\usepackage[latin1]{inputenc}
\usepackage{amsmath}
\usepackage{amssymb}

\begin{document}


\title{Lagrangian for Frenkel electron and position's non-commutativity due to spin.}

\author{Alexei A. Deriglazov}
\email{alexei.deriglazov@ufjf.edu.br}
\affiliation{%
Depto. de Matem\'atica, ICE,\\
Universidade Federal de Juiz de Fora, MG, Brasil\\ and\\
Laboratory of Mathematical
          Physics, Tomsk Polytechnic University,\\ 634050 Tomsk, Lenin Ave. 30, Russian Federation
}%

\author{Andrey M. Pupasov-Maksimov}
 \email{pupasov.maksimov@ufjf.edu.br}
\affiliation{%
Depto. de Matem\'atica, ICE,\\
Universidade Federal de Juiz de Fora, MG, Brasil}

\date{\today}

\begin{abstract}
We construct relativistic spinning-particle Lagrangian where spin is considered as a
composite quantity constructed on the base of non-Grassmann vector-like variable. The variational problem guarantees
both fixed value of spin and Frenkel condition on spin-tensor. The Frenkel condition
inevitably leads to relativistic corrections of the Poisson algebra of position variables: their classical brackets became
noncommutative. 
We construct the relativistic quantum mechanics in the canonical formalism (in
the physical-time parametrization) and in the covariant formalism (in an arbitrary parametrization). We show how state-vectors and
operators of the covariant formulation can be used to compute mean values of physical operators in the canonical formalism, thus
proving its relativistic covariance. 
We establish 
relations between Frenkel electron and positive-energy sector of Dirac equation.
Various candidates for position and spin operators of an
electron acquire clear meaning and interpretation in the 
Lagrangian model of Frenkel electron.
Our results
argue in favour of Pryce's (d)-type operators as
spin and position operators of Dirac theory.  This implies that effects {of non-commutativity could be expected already at the}
Compton wave length. We also present the manifestly
covariant form of spin and position operators of Dirac equation.
\end{abstract}

\maketitle


\section{Introduction and outlook \label{intro}}

Quantum description of spin is based on Dirac equation, whereas the most popular classical equations of electron have
been formulated by Frenkel \cite{Frenkel, Frenkel2} and Bargmann, Michel and Telegdi (F-BMT) \cite{BMT}. They almost
exactly reproduce spin dynamics of polarized beams in uniform fields, and agrees with the calculations based on Dirac
theory. Hence we expect that these models might be proper classical analog for the Dirac theory. The variational
formulation for F-BMT equations represents rather non trivial problem \cite{corben:1968, hanson1974relativistic,
berezin:1977, GG1, bt, bbra, grassberger1978,  cognola1981lagrangian, AAD5, AAD6, Alexei} (note that one needs a
Hamiltonian to study, for instance, Zeeman effect). In this work we continue systematic analysis of these equations
started in \cite{ Alexei}. We develop their Lagrangian formulation considering spin as a composite quantity (inner angular
momentum) constructed from non-Grass-\\
mann vector-like variable and its conjugated momentum \cite{grassberger1978,
cognola1981lagrangian, AAD5, AAD6, Alexei, deriglazov-ns:2010, AAD3, AAD4}.

Non relativistic spinning particle with reasonable properties can be constructed \cite{deriglazov-ns:2010, DPM1}
starting from singular Lagrangian which implies the following Dirac's constraints
\begin{eqnarray}\label{intr.1}
{\boldsymbol{\pi}}^2-a_3=0, \quad {\boldsymbol{\omega}}^2-a_4=0, \quad {\boldsymbol{\omega}}{\boldsymbol{\pi}}=0\,,
\end{eqnarray}
where $\quad a_3=\frac{3\hbar^2}{4a_4}$,
while relativistic form of these constraints read
\begin{eqnarray}\label{intr.2}
T_3=\pi^2-a_3=0\,, \quad T_4=\omega^2-a_4=0\,, \quad
T_5=\omega\pi=0\,, \\
T_6=   p\omega=0\,, \qquad T_7=p\pi=0\,.\label{intr.3}
\end{eqnarray}
Besides, we have the standard mass-shell constraint in
position sector, $T_1=p^2+(mc)^2=0$.
We denoted the basic variables of spin by $\omega^\mu=(\omega^0, {\boldsymbol\omega})$,
${\boldsymbol{\omega}}=(\omega_1, \omega_2, \omega_3)$, then $\omega\pi=-\omega^0\pi^0+{\boldsymbol\omega\boldsymbol\pi}$ and so
on. $\pi^\mu$ and $p^\mu$ are conjugate momenta for $\omega^\mu$ and the position $x^\mu$.

Since the constraints are written for the phase-space variables, it is easy to construct the corresponding action
functional in Hamiltonian formulation. We simply take $L_H=p\dot x+\pi\dot\omega-H$, with Hamiltonian in the form of linear
combination the constraints $T_i$ multiplied by auxiliary variables $g_i$, $i=1, 3, 4, 5, 6, 7$. The Hamiltonian action
with six auxiliary variables admits interaction with an arbitrary electromagnetic field and gives unified variational
formulation of both Frenkel and BMT equations, see \cite{Alexei}. In section
\ref{sec:labmt} we develop Lagrangian formulation of these equations. Excluding conjugate momenta from $L_H$, we obtain the
Lagrangian action. Further, excluding the auxiliary variables, one after another, we obtain various equivalent
formulations of the model. We shortly discuss all them, as they will be useful when we switch on interaction with
external fields \cite{DPW2, DPM2}. At the end, we get the ``minimal'' formulation without auxiliary variables. This reads
\begin{eqnarray}\label{intr.5}
S=\int\! d\tau\! \left(\sqrt{a_3\dot\omega N\dot\omega}
 -mc\sqrt{-\dot xN\dot x}-\frac{g_4}{2}(\omega^2-a_4)\right),
\end{eqnarray}
where $N^{\mu\nu}\equiv\eta^{\mu\nu}-\frac{\omega^\mu\omega^\nu}{\omega^2}$ is projector on the plane transverse to the
direction of $\omega^\mu$. The last term in (\ref{intr.5}) represents velocity-independent constraint which is well
known from classical mechanics. So, we might follow the classical-mechanics prescription to exclude the $g_4$ as well.
But this would lead to lose of manifest relativistic invariance of the formalism. The action is written in a
parametrization $\tau$ which obeys
\begin{eqnarray}\label{intr.6}
\frac{dt}{d\tau}> 0, \quad \mbox{this implies}\quad g_1(\tau)>0, \quad p^0>0.
\end{eqnarray}
To explain this restriction, we note that in absence of spin we expect an action of a spinless particle.  Switching off
the spin variables $\omega^\mu$ from Eq. (\ref{intr.5}), we obtain
$L=-mc\sqrt{-\dot x^2}$.
Let us compare this with spinless particle interacting with electromagnetic field. In terms of physical variables ${\bf
x}(t)$ this reads $L=-mc\sqrt{c^2-\dot{\bf x}^2}+eA_0+\frac{e}{c}{\bf A}\dot{\bf x}$. If we restrict ourselves to the
class of increasing parameterizations of the world-line, this reads $L=-mc\sqrt{-\dot x^2}+\frac{e}{c}A\dot x$, in
correspondence with spinless limit of (\ref{intr.5}).

Assuming $\frac{dt}{d\tau}<0$ we arrive at another Lagrangian, $L=mc\sqrt{-\dot x^2}+\frac{e}{c}A_\mu\dot x^\mu$. So a
variational formulation with both positive and negative parameterizations would describe simultaneously two classical
theories. In quantum theory they correspond to positive and negative energy solutions of Klein-Gordon equation
\cite{gitmantyutin1998}.

In \cite{DPM1} we discussed the geometry behind the constraints (\ref{intr.1})-(\ref{intr.3}). The phase-space surface
(\ref{intr.1}) can be identified with group manifold $SO(3)$. It has natural structure of fiber bundle with the base
being a two-dimensional sphere, thus providing a connection with the 
approach of Souriau \cite{souriau1970structure,souriau1997structure}.
Components of non relativistic spin-vector are defined by
$S_i=\epsilon_{ijk}\omega_j\pi_k$. At the end, they turn out to be functions of coordinates which parameterize the
base. The set (\ref{intr.2}), (\ref{intr.3}) is just a Lorentz-covariant form of the constraints (\ref{intr.1}). In the
covariant formulation, $S_i$ is included into the antisymmetric spin-tensor $J^{\mu\nu}=2\omega^{[\mu}\pi^{\nu]}$
according to the Frenkel rule, $J^{ij}=2\epsilon^{ijk}S^k$.

In the dynamical theory, these constraints can be interpreted as follows.
First, spin-sector constraints (\ref{intr.2}) fix the value of spin,
$J^{\mu\nu}J_{\mu\nu}=6\hbar^2$. As in the rest frame we have ${\bf
S}^2=\frac{1}{8}J^{\mu\nu}J_{\mu\nu}=\frac{3\hbar^2}{4}$, this implies the right value of three-dimensional spin, as
well as the right number of spin degrees of freedom.

Second, the first-class constraint $\pi^2-a_3=0$ provides additional local symmetry (spin-plane symmetry) of variational problem.
The spin-plane symmetry has clear geometric interpretation as transformations of structure group of the fiber bundle
acting independently at each instance of time. They rotate the pair ${\omega^\mu}, {\pi^\mu}$ in the plane formed by
these vectors. In contrast, $J^{\mu\nu}$ turns out to be invariant under the symmetry. Hence the spin-plane symmetry
determines physical sector of the spinning particle: the basic variable $\omega^\mu$ is gauge non-invariant, so does
not represent an observable quantity, while $J^{\mu\nu}$ does.

Reparametrization symmetry is known to be crucial for Lorentz-covariant description of a spinless particle. The
spin-plane symmetry, as it determines physical sector, turns out to be crucial for description of a spinning particle.
We point out that this appears already in non relativistic model (${\boldsymbol{\pi}}^2-a_3=0$ represents the
first-class constraint in the set (\ref{intr.1})). The local-symmetry group of minimal action will be discussed in some
details in subsection \ref{subsec:locs}. Curious property here is that the standard reparametrization symmetry turns
out to be combination of two independent local symmetries.

Equations (\ref{intr.3}) guarantee the Frenkel-type condition $J^{\mu\nu}p_\nu=0$. They form a pair of second-class
constraints which involve both spin-sector and position-sector variables. This leads to new properties as compared with
non relativistic formulation. The second-class constraints must be taken into account by transition from Poisson to
Dirac bracket. As the constraints involve conjugate momenta $p^\mu$ for $x^\mu$, this leads to nonvanishing Dirac
brackets for the position variables
\begin{eqnarray}\label{intr.4}
\{ x^\mu , x^\nu\}_D =-\frac{J^{\mu\nu}}{2p^2}.
\end{eqnarray}
We can pass from the parametric $x^\mu(\tau)$ to physical variables $x^i(t)$. They also obey a noncommutative algebra,
see Eq. (\ref{pha.15}) below. We remind that in a theory with second-class constraints one can find special coordinates
on the constraints surface with canonical (that is Poisson) bracket, see (\ref{pha.20.1}). Functions of special
coordinates are candidates for observable quantities. The Dirac bracket (more exactly, its nondegenerated part) is just
the canonical bracket rewritten in terms of initial coordinates \cite{gitman1990quantization}. For the present case,
namely the initial coordinates (they are $x^i(t)$), are of physical interest\footnote{In the interacting theory namely
the initial coordinates obey the F-BMT equations.}, as they represent the position of a particle. So, while there are
special coordinates with canonical symplectic structure, the physically interesting coordinates obey the
non-commutative algebra.

In the result, the position space is endowed, in a natural way, with noncommutative structure by
accounting spin degrees of freedom. Relations between spin and non-commutativity 
appeared already in the work of Matthisson \cite{mathisson1937zitternde,horvathy2003mathisson}. 
It is known that dynamical systems with second-class constraints
allow to incorporate noncommutative geometry into the framework of classical and quantum theory
\cite{hanson1974relativistic, Ghosh_94.1, Ghosh_94.2,deriglazov2010classical,Gomes2010spin-non-commutativity, Ferrari:2012bv, AAD10, AAD11}. Our model represents an example of situation when
physically interesting noncommutative particle (\ref{intr.4}) emerges in this way. For the case, the ``parameter of
non-commutativity'' is proportional to spin-tensor (spin non-commutativity imposed by hands in quantum theory is considered 
in \cite{Gomes2010spin-non-commutativity,Ferrari:2012bv}).

We point out that non relativistic model (\ref{intr.1}) implies canonical algebra of position operators, see
\cite{deriglazov-ns:2010, DPM1}. So the deformation (\ref{intr.4}) arises as a relativistic correction induced by spin
of the particle.

While the emergence of noncommutative structure in a classical theory is nothing more than a mathematical game, this
became crucial in quantum theory. Quantization of a theory with second-class constraints on the base of Poisson bracket
is not consistent, and we are forced to look for quantum realization of Dirac brackets. Instead of the standard
quantization rule of the position, $x\rightarrow\hat x=x$, we need to set $x\rightarrow\hat x=x+\hat\delta$ with some
operator $\hat\delta$ which provides the desired algebra (\ref{intr.4}). This leads to interesting consequences
concerning the relation between classical and quantum theories, which we start to discuss in this work.

A natural way to construct quantum observables is based on the correspondence principle between classical and quantum
descriptions. However, this straightforward approach is mostly restricted to simple models like non-relativistic point
particle. Elementary particles with spin were initially studied from the quantum perspective, because systematically
constructed classical models of spinning particle were non known. Construction of quantum observables for an electron
involves the analysis of Dirac equation and the representation theory of Lorentz group. Newton and Wigner found
possible position operator, $\hat{\bf x}_{NW}$, by the analysis of localized states in relativistic theory
\cite{newton1949localized}. Foldy and Wouthuysen invented a convenient representation for the Dirac equation
\cite{foldy:1978}. In this representation Newton-Wigner position operator simply becomes the multiplication operator,
$\hat{\bf x}_{NW}={\bf x}$. Pryce noticed that notion of center-of-mass in relativistic theory is not unique
\cite{pryce1948mass}, and suggested the list of possible operators. The Pryce center-of-mass (e) has commuting
components and coincides with the Newton-Wigner position operator, while the Pryce center-of-mass (d) is defined as a
covariant object though it has non-commutative components.

Notion of position observables in the theory of Dirac equation \cite{fleming1965covariant, bunge:69, kalnay:69} is in
close relation with the notion of relativistic spin. Current interest to covariant spin operators is related with a
broad range of physical problems concerning consistent definition of relativistic spin operator and Lorentz-covariant
spin density matrix in quantum information theory
\cite{chicone2005relativistic,singh2007cov-quantum-spin,slad:2010,KimSon2005,Caban2005spin-density-matrix,Jordan2006,Czachor1997,Landulfo2009}.
Consideration of {\it Zitterbewegung} \cite{cserti2006unified-Zitterbewegung} and spin currents \cite{Awschalom2002} in
condensed matter studies involves Heisenberg equations for position and spin observables. Precession of spin in
gravitational fields gives a useful tools to test general relativity \cite{Obukhov2009}. Surprisingly, coupling of spin
to gravitational fields may be important already in the acceleration experiments due to so-called spin-rotation
coupling \cite{Lambiase2013spin-rotation-coupling}. In these applications a better understanding of spinning particle
at the classical level may be very useful.

There are a lot of operators proposed for the position and spin of relativistic electron, see
\cite{pryce1948mass,newton1949localized,foldy:1978,feynman1961,corben:1968,Czachor1997}. Which one is a conventional
position (spin) operator? Widely assumed as the best candidate is the pair of Foldy-Wouthuysen ($\sim$ Newton-Wigner
$\sim$ Pryce (e)) mean position and spin operators. Components of the mean-position operator commutes with each other,
spin obeys $so(3)$ algebra. However, they do not represent Lorentz-covariant quantities.

To clarify these long-standing questions, in sections \ref{sec:min-action-physical}, \ref{sec:min-action-hamilton} and
\ref{sec:RQM} we construct relativistic quantum mechanics of F-BMT electron. In section \ref{sec:min-action-physical},
quantizing our Lagrangian in physical-time parametrization, we obtain the operators corresponding to classical position
and spin of our model. Our results argue in favor of covariant Pryce (d) position and spin operators\footnote{Pryce
(e)-operators corresponds to the special variables mentioned above, see subsection \ref{sec:min-action-physical_2}.}.
This implies that effects of non-commutativity could be presented at the Compton wave length, in contrast to
conventional expectations \cite{Kempf1995} of non-commutativity at Planck length.

In section \ref{sec:min-action-hamilton}, we construct Hamiltonian formulation in the covariant form (in an arbitrary
parametrization). The constraints $p^2+(mc)^2=0$ and $S^2=\frac{3\hbar^2}{4}$ appeared in classical model can be
identified with Casimir operators of Poincare group. That is the spin one-half representation of Poincare group
represents a natural quantum realization of our model. According to Wigner \cite{wigner1939unitary, Bargmann1948, Weinberg_2009}, this
is given by Hilbert space of solutions to two-component Klein-Gordon (KG) equation. Two-component KG field has been considered by
Feynman and Gell-Mann \cite{feynman1958fermi-interaction} to describe weak interaction of spin one-half particle in quantum field
theory, and by Brown \cite{brown1958two-component-fermi} as a starting point for QED. In contrast to KG equation for a scalar field, the
two-component KG equation admits the covariant positively defined conserved current
\begin{eqnarray}\label{intr.7}
I^\mu=\frac{1}{(mc)^2}(\bar\sigma\hat p\psi)^\dagger\sigma^\mu(\bar\sigma\hat p\psi)-\psi^\dagger\bar\sigma^\mu\psi\,,
\end{eqnarray}
which can be used to construct a relativistic quantum mechanics of this equation. This is done in subsection
\ref{subsec:KG}, then in subsection \ref{subsec:QM-BMT-Dirac-equiv} we show its equivalence with quantum mechanics of
Dirac equation. Taking into account the condition (\ref{intr.6}), we conclude that F-BMT electron corresponds to
positive-energy sector of the KG quantum mechanics, see subsection \ref{covop}. In subsection \ref{covinv}, we establish
the correspondence between canonical and covariant formulations of F-BMT electron, thus proving relativistic invariance
of the physical-time formalism of subsection \ref{sec:min-action-physical_2}. In particular, we find the manifestly-covariant operators
\begin{eqnarray}\label{cq13}
\hat x_{rp}^\mu &=& x^\mu+\frac{1}{2\hat p^2}(\sigma \hat p)^\mu\,,\\ 
\hat j^{\mu\nu} &=& \sigma^{\mu\nu}+\frac{\hat p^\mu(\sigma \hat p)^\nu-\hat p^\nu(\sigma \hat p)^\mu}{\hat p^2}\,,
\end{eqnarray}
and show how they can be used to compute mean values of the physical (that is Pryce (d)) operators of position and
spin. In other words, they represent manifestly-covariant form of Pryce (d)-operators.

Using the equivalence between KG and Dirac quantum mechanics, we then found the form of these operators on space of
Dirac spinors. They also can be used to compute position and spin of the Frenkel electron, see subsection \ref{covdir}.

\section{Search for Lagrangian \label{sec:labmt}}
\subsection{Variational problem with auxiliary variables \label{subsec:VariationalP-aux-vars}}
To start with, we take the Hamiltonian action \cite{Alexei}
\begin{eqnarray}\label{1.4}
S_H=\int d\tau \left(  p_\mu\dot x^\mu +\pi_\mu\dot\omega^\mu+\pi_{gi}\dot g_i-H\right) ,
\end{eqnarray}
\begin{eqnarray}\label{1.5}
\nonumber H &=& \frac{g_1}{2}(p^2+m^2c^2)+\frac{g_3}{2}(\pi^2-a_3)+\frac{g_4}{2}(\omega^2-a_4)+\\
&\,& g_5(\omega\pi)+g_6(p\omega)+g_7(p\pi)+\lambda_{gi}\pi_{gi}\,. \label{1.5}
\end{eqnarray}
Here $\pi_{gi}$ are conjugate momenta for the auxiliary variables $g_i$. We have denoted by $\lambda_{gi}$ the
Lagrangian multipliers for the primary constraints $\pi_{gi}=0$.
%
%
Variation of the action with respect to $\lambda_{gi}$ gives the equations $\pi_{gi}=0$, this implies $\dot\pi_{gi}=0$.
Using this in the equations $\frac{\delta S_H}{\delta g_i}=0$ we obtain \footnote{$\omega^\mu$ obeys the Hamiltonian
equation $\dot\omega^\mu=g_3\pi^\mu$. Together with $\pi^2>0$, this implies $\dot\omega^2>0$.} the desired constraints
(\ref{intr.2}) and (\ref{intr.3}). Our model is manifestly Poincare-invariant. The auxiliary variables $g_i$, are
scalars under the Poincare transformations. The remaining variables transform according to the rule
\begin{eqnarray}\label{1.5.1}
x'^{\mu} &=& \Lambda^\mu{}_\nu x^\nu+a^\mu, \qquad p'^\mu=\Lambda^\mu{}_\nu p^\nu, \\
\omega'^{\mu} &=& \Lambda^\mu{}_\nu\omega^\nu, \qquad \pi'^{\mu}=\Lambda^\mu{}_\nu\pi^\nu.
\end{eqnarray}
Local symmetries form two-parametric group of transformations. It is composed by the standard reparameterizations
\begin{eqnarray}\label{1.5.2}
\delta x^\mu &=& \alpha\dot x^\mu, \quad \delta p^\mu=\alpha\dot p^\mu,\\
\delta\omega^\mu &=& \alpha\dot\omega^\mu, \quad
\delta\pi^\mu=\alpha\dot\pi^\mu,\\
 \delta g_i &=& (\alpha g_i)\dot{}, \qquad \delta\lambda_{g_i}=(\delta g_i)\dot{}.
\end{eqnarray}
as well as by spin-plane transformations with the parameter $\beta(\tau)$:
\begin{eqnarray}\label{3.5.1}
\delta\omega^\mu &=& s\beta\pi^\mu, \quad \delta\pi^\mu=-\frac{1}{s}\beta\omega^\mu,\\
\label{3.5.2}
\delta g_3 &=& s\dot\beta-2sg_5\beta, \quad \delta g_4=\frac{1}{s}\dot\beta+2\frac{1}{s}g_5\beta, \cr 
\delta g_6 &=& \frac{1}{s}\beta g_7, \quad \delta g_7=-s\beta g_6, \cr 
\delta g_5 &=& \frac{1}{s}\beta g_3-s\beta g_4, \quad
\delta\lambda_{gi}=(\delta g_i)\dot{}.
\end{eqnarray}
We have denoted $s\equiv\sqrt{\frac{a_4}{a_3}}$. Eq. (\ref{3.5.1}) represents infinitesimal form of the structure-group
transformations of the spin-fiber bundle \cite{DPM1}.

The coordinates $x^\mu$, Frenkel spin-tensor $J^{\mu\nu}$ and BMT vector $s_{BMT}^\mu$
\begin{eqnarray}\label{ha.10.1}
J^{\mu\nu}(\tau) &=& 2(\omega^\mu\pi^\nu-\omega^\nu\pi^\mu)\,, \\
s_{BMT}^\mu(\tau) &\equiv& \frac{1}{4\sqrt{-p^2}}\epsilon^{\mu\nu\alpha\beta}p_\nu J_{\alpha\beta}\,,
\end{eqnarray}
are $\beta$\,-invariant quantities. For their properties see Appendix 1. 
Note that the spacial components, $s_{BMT}^i$, coincide with Frenkel spin
\begin{eqnarray}\label{spin}
S^i=\frac{1}{4}\epsilon^{ijk}J_{jk}\,,
\end{eqnarray}
only in the rest frame. Both transform as a vector under spacial rotations, but have different transformation laws
under Lorentz boost. In an arbitrary frame they are related by
\begin{eqnarray}\label{spin.1}
S^i=\frac{p^0}{\sqrt{-p^2}}\left(\delta_{ij}-\frac{p_ip_j}{(p^0)^2}\right)s_{BMT}^j\,,
\end{eqnarray}
Where this does not lead to misunderstanding, we denote $s_{BMT}^\mu$ as $s^\mu$.

Lagrangian of a given Hamiltonian theory with constraints can be restored within the known procedure
\cite{gitman1990quantization,deriglazov2010classical}. For the present case, it is sufficiently to solve Hamiltonian
equations of motion for $x^\mu$ and $\omega^\mu$ with respect to $p^\mu$ and $\pi^\mu$, and substitute them into the
Hamiltonian action (\ref{1.4}). Let us do this for more general Hamiltonian action, obtaining closed formula which will
be repeatedly used below.

Consider mechanics with the configuration-space variables $Q^a(\tau)$, $g_i(\tau)$, and with the Lagrangian action
\begin{eqnarray}\label{s.1}
S=\frac12\int d\tau \left( G_{ab}DQ^aDQ^b-K_{ab}Q^aQ^b-M\right) .
\end{eqnarray}
We have denoted $DQ^a\equiv\dot Q^a-H^a{}_bQ^b$, and $G(g, Q)$, $K(g, Q)$, $H(g, Q)$, and $M(g)$ are some functions of
the indicated variables. Let us construct the Hamiltonian action functional of this theory. Denoting the conjugate
momenta as $P_a$, $\pi_{gi}$, the equations for $P_a$ can be solved
\begin{eqnarray}\label{s.2}
P_a=\frac{\partial L}{\partial\dot Q^a}=G_{ab}DQ^b, ~ \Rightarrow ~ \dot Q^a=\tilde G^{ab}P_b+H^a{}_bQ^b,
\end{eqnarray}
where $\tilde G^{ab}$ is the inverse matrix of $G_{ab}$. Equations for the remaining momenta turn out to be the primary
constraints, $\pi_{gi}=0$. Then the Hamiltonian action reads
\begin{eqnarray}\label{s.3}
S_H=\int d\tau \left(P_a\dot Q^a+\pi_{gi}\dot g_i-H\right),
\end{eqnarray}
\begin{eqnarray}\label{s.4}
H &=& \frac12\tilde G^{ab}P_aP_b+P_aH^a{}_bQ^b+\frac12K_{ab}Q^aQ^b+\frac12M+\cr
&\,& \lambda_{gi}\pi_{gi}.
\end{eqnarray}
Thus the Hamiltonian (\ref{s.3}) and the Lagrangian (\ref{s.1}) variational problems are equivalent. We point out that
choosing an appropriate set of auxiliary variables $g_i$, the action (\ref{s.1}) can be used to produce any desired
quadratic constraints of the variables $Q, P$.

Let us return to our problem (\ref{1.5}). Comparing the Hamiltonian of our interest (\ref{1.5}) with the expression
(\ref{s.4}), we define the "doublets" $Q^a=(x^\mu, \omega^\nu)$, $P_a=(p_\mu, \pi_\nu)$, as well as the matrices
\begin{eqnarray}\label{2.4}
\tilde G^{ab}=\left(
\begin{array}{cc}
g_1 & g_7\\
g_7 & g_3
\end{array}
\right), ~ H^a{}_b=\left(
\begin{array}{cc}
0 & g_6\\
0 & g_5
\end{array}
\right), ~ K_{ab}=\left(
\begin{array}{cc}
0 & 0\\
0 & g_4
\end{array}
\right), \nonumber
\end{eqnarray}
where $g_1=g_1\eta^{\mu\nu}$ and so on.  Besides, we take the "mass" term in the form $M=g_1m^2c^2-a_3g_3-a_4g_4$. With
this choice, the equation (\ref{s.4}) turns into our Hamiltonian (\ref{1.5}). So the corresponding Lagrangian action
reads from (\ref{s.1}) as follows
\begin{eqnarray}\nonumber
S&=&\int d\tau\frac{1}{2\det\tilde G}\left[g_3(Dx)^2-2g_7(DxD\omega)+g_1(D\omega)^2\right]-\\
&\,& \frac12g_1m^2c^2+\frac{1}{2}g_3a_3-\frac{1}{2}g_4(\omega^2-a_4).\label{2.5}
\end{eqnarray}
We have denoted
\begin{eqnarray}\label{s.5.1}
Dx^\mu=\dot x^\mu-g_6\omega^\mu, \qquad D\omega^\mu=\dot\omega^\mu-g_5\omega^\mu. \nonumber
\end{eqnarray}
Using the inverse matrix
\begin{eqnarray}\label{2.4a}
G_{ab}=\frac{1}{\det\tilde G}\left(
\begin{array}{cc}
g_3 & -g_7\\
-g_7 & g_1
\end{array}
\right), \quad \nonumber
\end{eqnarray}
the action can be written in the form
\begin{eqnarray}\label{2.5a}
\nonumber S=\int d\tau \left( \frac{1}{2}G_{ab}DQ^aDQ^b+\frac{a_3g_{11}-m^2c^2g_{22}}{2\det G}-\frac{1}{2}g_4(\omega^2-a_4)\right),
\end{eqnarray}
where $DQ^a=(Dx, D\omega)$.

\subsection{Variational problem without auxiliary variables \label{subsec:VariationalP-without-aux-vars}}
Eliminating the auxiliary variables one by one, we get various equivalent formulations of the model (\ref{2.5}). At the
end, we arrive at the Lagrangian action without auxiliary variables $g_i$.

First, we write equations for $g_5$ and $g_6$ following from (\ref{2.5}). They imply $(\omega D\omega)=0$ and $(\omega
Dx)=0$, then
\begin{eqnarray}\label{s.7}
g_5=\frac{(\dot\omega\omega)}{\omega^2}, \qquad g_6=\frac{(\dot x\omega)}{\omega^2}. \nonumber
\end{eqnarray}
We substitute the solution\footnote{There is no guarantee that this gives an equivalent variational problem, the
equivalence must be verified by direct computations. Fortunately, for our case the trick works well.} into the action
(\ref{2.5}), this reads
\begin{eqnarray}
S &=&\int d\tau\left( \frac{\left[g_3(\dot xN\dot x)-2g_7(\dot xN\dot\omega)+g_1(\dot\omega N\dot\omega)\right]}{2\det\tilde G}-
 \frac12g_1m^2c^2+\frac{1}{2}g_3a_3-\frac{1}{2}g_4(\omega^2-a_4)\right) .\label{s.6}
\end{eqnarray}
It has been denoted
\begin{equation}\label{s.9}
N^{\mu\nu}\equiv\eta^{\mu\nu}-\frac{\omega^\mu\omega^\nu}{\omega^2}\,, \quad \mbox{then} \quad N^{\mu\nu}\omega_\nu=0.
\end{equation}
Together with $\tilde N^{\mu\nu}\equiv\frac{\omega^\mu\omega^\nu}{\omega^2}$, this forms a pair of projectors
$N+\tilde N=1$, $N^2=N$, $\tilde N^2=\tilde N$, $N\tilde N=0$.
Any vector $V^\mu$ can be decomposed on the transverse and longitudinal parts with respect to $\omega^\mu$,
$V^\mu=V_{\perp}^\mu+V_{\parallel}^\mu$, where
$V_{\perp}^\mu=N^\mu{}_\nu V^\nu$, then $V_{\perp}^\mu\omega_\mu=0$;
and
$V_{\parallel}^\mu=\tilde N^\mu{}_\nu V^\nu=\frac{(\omega V)}{\omega^2}\omega^\mu\sim\omega^\mu$.
Further, in the action (\ref{s.6}) we put $g_7=0$
\begin{eqnarray}
S=\int d\tau \left( \frac{1}{2g_1}(\dot xN\dot x)-\frac12g_1m^2c^2+
\frac{1}{2g_3}(\dot\omega
N\dot\omega)+\frac12a_3g_3-\frac12g_4(\omega^2-a_4)\right).\label{s.8}
\end{eqnarray}
This does not alter the dynamical equations, whereas the constraint $\omega\pi=0$ appears as the third-stage
constraint.

The first two terms in Eq. (\ref{s.8}) (as well as the third and the fourth terms) have the structure similar to that
of spinless particle, $\frac{1}{2e}\dot x^2-\frac{em^2c^2}{2}$. It is well known, that for the case we can substitute
equations of motion for $e$ back into the Lagrangian, this leads to an equivalent variational problem. So, we solve the
equation for $g_3$, $g_3=\sqrt{\frac{\dot\omega N\dot\omega}{a_3}}$, and substitute this back into (\ref{s.8}), this
gives
\begin{eqnarray}\label{s.14}
S=\int d\tau \left( \frac{1}{2g_1}(\dot xN\dot x)-\frac12g_1m^2c^2+\sqrt{a_3}\sqrt{\dot\omega N\dot\omega}-\frac{1}{2}g_4(\omega^2-a_4)\right).
\end{eqnarray}
Analogously, we solve the equation for $g_1$, $g_1=\frac{\sqrt{-\dot x N\dot x}}{mc}$ and substitute this into
(\ref{s.14}), this gives the "minimal" action
\begin{eqnarray}\label{s.15}
S=\! \int\! d\tau \!\left[\sqrt{a_3\dot\omega N\dot\omega}
 -mc\sqrt{-\dot xN\dot x}-\frac{g_4}{2}(\omega^2-a_4)\right].
\end{eqnarray}
This depends only on transverse parts of the velocities $\dot x^\mu$ and $\dot\omega^\mu$. The second term from
(\ref{s.15}) appeared as a Lagrangian of the particle \cite{dullemond1983, Tarakanov2006} inspired by Bag model
\cite{Chodos1974} in hadron physics.

\subsection{Local symmetries of minimal action \label{subsec:locs}}
Our model is invariant under two local symmetries. For the initial formulation (\ref{1.4}) they have been written in
Eqs. (\ref{1.5.1}) and (\ref{1.5.2}). Let us see how they look for the minimal action. This is invariant under
reparametrization of the lines $x^\mu(\tau)$ and $\omega^\mu(\tau)$ supplemented by proper transformation of the
auxiliary variable $g_4(\tau)$. We use the projectors $N$ and $\tilde N$ to decompose an infinitesimal reparametrization as
follows:
\begin{eqnarray}\label{lo.1}
\delta x^\mu &=&\alpha\dot x^\mu=\alpha\tilde N\dot x^\mu+\alpha N\dot x^\mu, \\
\delta\omega^\mu &=& \alpha\dot\omega^\mu=\alpha N\dot\omega^\mu+\alpha\tilde N\dot\omega^\mu,\nonumber \\
\delta g_4 &=& (\alpha
g_4)\dot{}=\left(\frac{\alpha\sqrt{a_3\dot\omega N\dot\omega}}{\omega^2}\right)\dot{}+\left(\alpha
g_4-\alpha\frac{\sqrt{a_3\dot\omega N\dot\omega}}{\omega^2}\right)\dot{}.\nonumber
\end{eqnarray}
Our observation is that each projection
\begin{eqnarray}\label{lo.2}
\delta_\beta x^\mu &=& \beta\tilde N\dot x^\mu, \qquad \delta_\beta\omega^\mu=\beta N\dot\omega^\mu, \cr 
\delta_\beta
g_4 &=& \left(\beta\frac{\sqrt{a_3}\sqrt{\dot\omega N\dot\omega}}{\omega^2}\right)\dot{}.
\end{eqnarray}
\begin{eqnarray}\label{lo.3}
\delta_\gamma x^\mu &=& \gamma N\dot x^\mu, \qquad \delta_\gamma\omega^\mu=\gamma\tilde N\dot\omega^\mu, \cr
\delta_\gamma g_4 &=& \left(\gamma g_4-\gamma\frac{\sqrt{a_3}\sqrt{\dot\omega N\dot\omega}}{\omega^2}\right)\dot{}.
\end{eqnarray}
separately turns out to be a symmetry of the minimal action. It can be verified using the intermediate expressions
\begin{eqnarray}\label{lo.4}
\delta_\beta\omega^2 &=& 0, \quad \delta_\beta\sqrt{-\dot xN\dot x}=0\,,
\cr 
\delta_\beta N^{\mu\nu}&=&-\frac{\beta}{\omega^2}((N\dot\omega)^\mu\omega^\nu+(\mu
\leftrightarrow\nu)), \cr 
\delta_\beta\sqrt{\dot\omega
N\dot\omega} &=& \left(\beta\sqrt{\dot\omega N\dot\omega}\right)\dot{}-(\omega^2)\dot{}\frac{\beta\sqrt{\dot\omega
N\dot\omega}}{2\omega^2}\,, \nonumber
\label{lo.5}\\
\delta_\gamma\omega^2 &=& \gamma(\omega^2)\dot{}, \quad \delta_\gamma N^{\mu\nu}=0, \cr 
\delta_\gamma\sqrt{-\dot
xN\dot x} &=& (\gamma\sqrt{-\dot x N\dot x})\dot{}, \cr 
\delta_\gamma\sqrt{\dot\omega
N\dot\omega} &=& (\omega^2)\dot{}\frac{\gamma\sqrt{\dot\omega N\dot\omega}}{2\omega^2}. \nonumber
\end{eqnarray}
Any pair among the transformations (\ref{lo.1})-(\ref{lo.3}) can be taken as independent symmetries of the minimal
action.

Let the functions $x(\tau), \omega(\tau), g_4(\tau)$ represent a solution to equations of motion. Then they obey
$(\omega\dot\omega)=(\omega\dot x)=0$ and $g_4=\frac{\sqrt{a_3}\sqrt{\dot\omega N\dot\omega}}{\omega^2}$. Using this
expressions, the transformations (\ref{lo.2}) and (\ref{lo.3}) acquire the form
\begin{eqnarray}\label{lo.6}
\delta_\beta x^\mu=0, \quad \delta_\beta\omega^\mu=\beta\dot\omega^\mu, \quad \delta_\beta g_4=(\beta g_4)\dot{}.
\end{eqnarray}
\begin{eqnarray}\label{lo.7}
\delta_\gamma x^\mu=\gamma\dot x^\mu, \quad \delta_\gamma\omega^\mu=0, \quad \delta_\gamma g_4=0.
\end{eqnarray}
Hence on true trajectories the symmetries have simple meaning. $\gamma$\,-transformations (\ref{lo.7}) represent
reparametrizations of the configuration-space trajectory $x^\mu$, whereas $\beta$\,-transformations (\ref{lo.6})
represent reparametrizations of the inner-space trajectory $\omega^\mu$. Their sum gives the standard reparametrization
transformation of the theory, Eq. (\ref{lo.1}).

\section{Minimal action in the physical-time parametrization  \label{sec:min-action-physical}}
\subsection{Position's non-commutativity due to spin \label{sec:min-action-physical_1}}
Using reparametrization invariance  of the Lagrangian (\ref{s.15}), we take physical time as the evolution parameter,
$\tau=t$. Now we work with physical dynamical variables $x^\mu=(ct, {\bf x}(t))$ and $\omega^\mu=(\omega^0(t),
{\boldsymbol\omega}(t))$ in the expression (\ref{s.15}). In this section the dot means derivative with respect to $t$,
$\dot x^\mu=(c, \frac{d{\bf x}}{dt})$ and so on. Let us construct Hamiltonian formulation of the model (\ref{s.15}).

Computing conjugate momenta, we obtain the primary constraint $\pi_{g4}=0$, and the expressions
\begin{eqnarray}\label{pha.3}
p_i=mc\frac{N\dot x^i}{\sqrt{-\dot xN\dot x}},
\end{eqnarray}
\begin{eqnarray}\label{pha.3.1}
\pi^\mu=\sqrt{a_3}\frac{N\dot\omega^\mu}{\sqrt{\dot\omega
N\dot\omega}}.
\end{eqnarray}
Comparing expressions for ${\bf p}^2$ and ${\bf p}{\boldsymbol\omega}$, after tedious computations we obtain the
equality which does not involve time-derivative, ${\bf p}^2+(mc)^2=(\frac{{\bf p}{\boldsymbol\omega}}{\omega^0})^2$.
Hence Eq. (\ref{pha.3}) implies the constraint
\begin{eqnarray}\label{pha.3.2}
-\sqrt{{\bf p}^2+(mc)^2}\omega^0+{\bf p}{\boldsymbol\omega}=0. \nonumber
\end{eqnarray}
This is analog of covariant constraint $p^\mu\omega_\mu=0$. Eq. (\ref{pha.3.1}) together with Eq. (\ref{s.9}) imply
more primary constraints
$\omega\pi=0$, $\pi^2-a_3=0$.
Computing the Hamiltonian, $P\dot Q-L+\lambda_a\Phi_a$, we obtain
\begin{eqnarray}\label{pha.6}
& H&= c\sqrt{{\bf p}^2+(mc)^2}+\lambda_3(\pi^2-a_3)+\frac{1}{2}g_4(\omega^2-a_4)+\cr
&\,& \lambda_5(\omega\pi)+\lambda_6(-\sqrt{{\bf p}^2+(mc)^2}\omega^0+{\bf p}{\boldsymbol\omega})+\lambda_4\pi_{g4}.
\end{eqnarray}
Preservation in time of the primary constraints implies the following chains of algebraic consequences:
\begin{eqnarray}\label{pha.7}
\pi_{g4} &=& 0,  ~ \Rightarrow ~ \omega^2-a_4=0, ~ \Rightarrow ~ \lambda_5=0. \cr 
(\omega\pi) &=& 0, ~ \Rightarrow ~ \lambda_3=\frac{a_4}{2a_3}g_4.   \cr
&-&\sqrt{{\bf p}^2+(mc)^2}\omega^0+{\bf p}{\boldsymbol\omega}=0,  ~ \Rightarrow ~\cr 
&-&\sqrt{{\bf p}^2+(mc)^2}\pi^0+{\bf p}{\boldsymbol\pi}=0, ~ \Rightarrow ~ \lambda_6=0. \nonumber
\end{eqnarray}
Three Lagrangian multipliers have been determined in the process, $\lambda_5=\lambda_6=0$ and
$\lambda_3=\frac{a_4}{2a_3}g_4$, whereas $\lambda_1$ and $\lambda_4$ remain arbitrary functions. For the latter use,
let us denote
\begin{eqnarray}\label{pha.7.1}
p^0\equiv\sqrt{(mc)^2+{\bf p}^2}\,, \qquad g^{\mu\nu}\equiv\eta^{\mu\nu}-\frac{p^{\mu}p^\nu}{p^2}\,.
\end{eqnarray}
Besides the constraints, the action implies the Hamiltonian equations
\begin{eqnarray}\label{pha.8}
\frac{dx^i}{dt}=c\frac{p^i}{p^0}, \quad \frac{dp^i}{dt}=0,
\end{eqnarray}
\begin{eqnarray}\label{pha.9}
\dot g_4=\lambda_{4}, \qquad \dot\pi_{g4}=0;
\end{eqnarray}
\begin{eqnarray}\label{pha.10}
\dot\omega^\mu=\frac{a_4}{a_3}g_4\pi^\mu, \qquad \dot\pi^\mu=-g_4\omega^\mu.
\end{eqnarray}
Equations (\ref{pha.8}) describe free-moving particle with the speed less then speed of light
\begin{eqnarray}\label{pha.11}
x^i=x^i_0+v^it, \quad v^i=c\frac{p^i}{\sqrt{(mc)^2+{\bf p}^2}}, \quad p^i=\mbox{const}.
\end{eqnarray}
The spin-sector variables have ambiguous evolution, because a general solution to (\ref{pha.10}) depends on an arbitrary
function $g_4$. So they do not represent the observable quantities. As candidates for the physical variables of
spin-sector, we can take either the Frenkel spin-tensor,
\begin{eqnarray}\label{pha.12}
\frac{dJ^{\mu\nu}}{dt}=0, \quad J^{\mu\nu}p_\nu=0, \quad J^2=6\hbar^2.
\end{eqnarray}
or, equivalently, BMT vector
\begin{eqnarray}\label{pha.14}
\frac{ds^\mu}{dt}=0, \quad s^\mu p_\nu=0, \quad s^2=\frac{3\hbar^2}{4}.
\end{eqnarray}
The constraints $\pi^2-a_3=0$ and $\pi_{g4}=0$ belong to first-class, other form the second-class set. To take the
latter into account, we construct the corresponding Dirac bracket. The non vanishing Dirac brackets are
\begin{eqnarray}\label{pha.15}
\{x^i,x^j\}_{D}&=&\frac{\epsilon^{ijk}s_k}{mcp^0}\,,\quad \{x^i,p^j\}_{D}=\delta^{ij}\,,\\
\{p^i,p^j\}_{D} &=& 0,\\
\label{cq3}
\{J^{\mu\nu},J^{\alpha\beta}\}_D &=& 2\left(g^{\alpha \left[\mu\right.}
J^{\left.\nu\right]\beta}-g^{\beta\left[\mu\right.} J^{\left.\nu\right]\alpha}\right)\,,\\
\label{xmJab-bmt-diracD}
\{x^\mu, J^{\alpha\beta}\}_{D} &=& \frac{1}{(mc)^2}\left(J^{\mu\left[\beta\right.}p^{\left.\alpha\right]}
-\frac{p^\mu}{p^0} J^{0\left[\beta\right.}p^{\left.\alpha\right]}\right) \,,\\
\label{pha.16}
\{s^i,s^j\}_D &=&\frac{p^0}{mc}\epsilon^{ijk}\left(s_k- \frac{({\bf{s}\,\bf{p}})p_k}{p_0^2}\right)\,,\\
\label{pha.17}
\{x^i,s^j\}_D &=& \left(s^i-\frac{({\bf{s}\, \bf{p}})p^i}{p_0^2}\right)\frac{p^j}{(mc)^2}\,,
\end{eqnarray}
where $p^0$ and $g^{\mu\nu}$ have been specified in (\ref{pha.7.1}). After transition to the Dirac brackets the
second-class constraints can be used as strong equalities. In particular, we can present $s^0$ in terms of independent
variables
\begin{eqnarray}\label{pha.18}
s^0=\frac{(\bf{s}\, \bf{p})}{\sqrt{{\bf p}^2+(mc)^2}}\,, \nonumber
\end{eqnarray}
and in the expression for Hamiltonian (\ref{pha.6}) only first and second terms survive. Besides, we omit the second term, as it does
not contribute into equations for spin-plane invariant variables. In the result, we obtain the physical Hamiltonian
\begin{eqnarray}\label{pha.19}
H_{ph}=c\sqrt{{\bf p}^2+(mc)^2}\,.
\end{eqnarray}
As it should be, the equations (\ref{pha.8}), (\ref{pha.12}) and (\ref{pha.14}) follow from physical Hamiltonian
with use the Dirac bracket, $\dot Q=\{ Q, H_{ph}\}_D$.

\subsection{Operators of physical observables: F-BMT electron chooses Pryce's (d)\,-type spin and position \label{sec:min-action-physical_2}}
Both operators (except $\hat p_i$) and abstract state-vectors of the physical-time formalism we denote by capital
letters, $\hat Q$, $\Psi(t, {\bf x})$. In order to quantize the model, classical Dirac-bracket algebra should be
realized by operators, $[\hat Q_1 , \hat Q_2]=i\hbar \left.\{ Q_1 , Q_2 \}_D\right|_{Q_i\rightarrow\hat Q_i}$. To start
with, we look for classical variables which have canonical Dirac brackets, thus simplifying the quantization procedure.
Consider the spin variables $\tilde{s}_j$ defined by the following transformation:
\[
\tilde{s}_j=\left(\delta_{jk}-\frac{p_jp_k}{p^0(p^0+mc)}\right)s^k\,,\]
\[
s_j=\left(\delta_{jk}+\frac{p_jp_k}{mc(p^0+mc)}\right)\tilde{s}^k\,.
\]
Vector $\bf{\tilde{s}}$ is nothing but the spin in the rest frame. Its components have the following Dirac brackets
\begin{eqnarray}\label{SmSn-modif-DiracBracketsD}
\{\tilde{s}^i,\tilde{s}^j\}_D &=&\epsilon^{ijk}\tilde{s}_k\,,\\
\{x^i,\tilde{s}^j\}_D &=&\frac{1}{mc(p^0+mc)}\left(\tilde{s}^ip^j-\delta^{ij}(\bf{p}\,\bf{\tilde{s}})\right)\,. \nonumber
\end{eqnarray}
The last equation together with the following Dirac bracket:
$\{\epsilon^{ikm}\tilde{s}_kp_m,\tilde{s}^j\}=\tilde{s}^ip^j-\delta^{ij}(\bf{p}\,\bf{\tilde{s}})$,
suggest to consider the variables
\begin{eqnarray}\label{pha.20}
\tilde{x}^j=x^j-\frac{1}{mc(p^0+mc)}\epsilon^{jkm}\tilde{s}_kp_m\,.
\end{eqnarray}
The canonical variables $\tilde{x}^j$, $p_i$ and $\tilde{S}^j$ have a simple algebra
\begin{eqnarray}\label{pha.20.1}
\{\tilde{x}^j,\tilde{x}^i\}_D &=& 0\,,\qquad \{\tilde{x}^i,p^j\}_D=\delta^{ij}\,,\\ 
\{\tilde{x}^j,\tilde{s}^i\}_D &=& 0\,,
\qquad \{\tilde{s}^i,\tilde{s}^j\}_D=\epsilon^{ijk}\tilde{s}_k\,.
\end{eqnarray}
Besides, the constraints (\ref{pha.14}) on $s^\mu$ imply ${\bf\tilde{s}}^2=\frac{3}{4}\hbar^2$. So the corresponding operators
$\hat{\tilde{S}}^j$ should realize an irreducible representation of $SO(3)$ with spin $s=1/2$. Quantization in terms of
these variables becomes straightforward. The Hilbert space consists from two-component functions $\Psi_a(t, {\bf x})$,
$a=1, 2$.
A realization of Dirac brackets algebra by operators has the standard form
\[
p_j\to \hat{p}_j=-i\hbar \partial_j\,,\]
\[ \tilde x^j\to\hat{\tilde{X}}^j=x^j\,,\]
\[ \tilde{s}_{BMT}^j\to
\hat{\tilde{S}}_{BMT}^j=\frac{\hbar}{2}\sigma^j\,.
\]
The conversion formulas between canonical and initial variables have no ordering ambiguities, so we immediately obtain
the operators corresponding to the physical position and spin of classical theory
\begin{equation}\label{electron-position-d-components-FW}
x^i ~  \rightarrow ~ \hat{X}^{i}=x^i-\frac{\hbar}{2mc(\hat p^0+mc)}\epsilon^{ijk}\hat{p}_j\sigma_k\,,
\end{equation}
\begin{equation}\label{pha.21}
\hat{J}^{0i}=-\frac{\hbar}{mc}\epsilon^{ijk}\hat p_j \sigma_k\,, 
\end{equation}
\begin{equation}\label{pha.22}
\hat J^{ij}=\frac{\hbar}{mc}\epsilon^{ijk}\left(\hat p^0\sigma_k-\frac{1}{(\hat
p^0+mc)}(\hat{\bf{p}}{\boldsymbol{\sigma}})\hat p_k\right) \,,
\end{equation}
\begin{equation}\label{spin.1}
\hat S^i=\frac{1}{4}\epsilon^{ijk}\hat J_{jk}=\frac{\hbar}{2mc}\left(-\hat p^0\sigma^i-\frac{1}{(\hat
p^0+mc)}(\hat{\bf{p}}{\boldsymbol{\sigma}})\hat p^i\right) \,.
\end{equation}
BMT operator reads
\begin{equation}\label{S0-operator}
\hat{S}_{BMT}^0=-\frac{\hbar}{2mc}({\hat{\bf{p}}\boldsymbol{\sigma}})\,,
\end{equation}
\begin{equation}\label{Sj-operator}
\hat{S}_{BMT}^j=\frac{\hbar}{2}\left(\sigma^j+\frac{1}{mc(\hat p^0+mc)}(\hat{\bf{p}}{\boldsymbol{\sigma}})\hat
p^j\right) \,.
\end{equation}
The energy operator (\ref{pha.19})
determines the evolution of a state-vector by the Schr\"{o}dinger equation
\begin{equation}\label{schrodinger-eq-bmt}
i\hbar\frac{d\Psi}{dt}=c\sqrt{\hat{{\bf p}}^{\,2}+(mc)^2}\Psi\,,
\end{equation}
as well the evolution of operators by Heisenberg equations. The scalar product can be defined as follows
\begin{equation}\label{fw-scalar-product}
\langle \Psi,\Phi \rangle=\int d^3x\Psi^\dagger \Phi\,.
\end{equation}
By construction, the abstract vector $\Psi(t, {\bf x})$ of Hilbert space can be identified with amplitude of
probability density of canonical coordinate $\tilde x^i$. Since our position operators $\hat x^i$ are noncommutative, the
issue of a wave function requires special discussion which we postpone for the future.
\begin{table*}
\begin{center}
\caption{Position/spin operators for the relativistic electron \cite{pryce1948mass}} $\beta=\left(
\begin{array}{cc}
1& 0\\
0& -1
\end{array}
\right)$, $\alpha^i=\left(
\begin{array}{cc}
0& \sigma^i\\
\sigma^i& 0
\end{array}
\right)$, $\Sigma^i=\left(
\begin{array}{cc}
\sigma^i& 0\\
0& \sigma^i
\end{array}
\right)$. \label{tabular:Pryce-cm} \vspace{2mm}
\begin{center}
\begin{tabular*}{\textwidth}{c|c|c|c}
{}  & Dirac representation, $i\hbar\partial_t\Psi_D=c(\alpha^ip_i+mc\beta)\Psi_D$   & F-W representation, $i\hbar\partial_t\Psi=c\beta\hat p^0\Psi$
&Classical model \\
\hline \hline $\hat {X}_{P(d)}^j$           &
$x^j+\frac{i\hbar}{2mc}\beta\left(\alpha^j-\frac{\alpha^k\hat{p}_k\hat{p}^j}{(\hat p^0)^2}\right)$    &
$x^j-\frac{\hbar\epsilon^{jkm}\hat{p}_k \Sigma_m}{2mc(\hat p^0+mc)}$ &position $x^j$
\\
$\hat{S}_{P(d)}^{j}$       &  $\frac{1}{2m^2c^2}\left(m^2c^2 \Sigma^j -imc\beta \epsilon^{jkl}\alpha_k\hat{p}_{l}
 \right)$ & $\frac{\hbar}{2mc}\beta\left(\hat p^0 \Sigma^j-\frac{\hat{p}^k\Sigma_k \hat p^j}{(\hat p^0+mc)}\right)$ &Frenkel
 spin $S^j$
\\
\hline $\hat{X}_{P(e)}^j=\hat{x}_{FW}^j$  & $x^j+\frac{\hbar}{2\hat p^0}\left(i \beta\alpha^j +\frac{\epsilon^{jkm}\hat{p}_k \Sigma_m}{\hat
p^0+mc} -\frac{i\beta \alpha^k\hat{p}_k\hat{p}^j}{\hat p^0(\hat p^0+mc)}\right) $ &
$x^j$ &$\tilde x^j$
\\
$\hat{S}_{P(e)}^{j}=\hat{S}_{FW}^{j}$       & $\frac{\hbar}{2\hat p^0}\left(mc \Sigma^j -im\beta
\epsilon^{jkl}\alpha_k\hat{p}_{l}+\frac{\Sigma^k\hat{p}_k\hat{p}^j}{\hat p^0+mc}
 \right)$ & $\frac{\hbar}{2}\Sigma^j $ &$\tilde s^j$
\\
\hline $\hat X_{P(c)}^j$           & $x^j+\frac{\hbar}{2(\hat p^0)^2}\left(\epsilon^{jkm}\hat{p}_k \Sigma_m +i mc
\beta\alpha^j \right)$    & $x^j+\frac{\hbar\epsilon^{jkm}\hat{p}_k \Sigma_m}{2\hat p^0(\hat p^0+mc)}$
\\
$\hat{S}_{P(c)}^{j}$       & $\frac{\hbar}{2(\hat p^0)^2}\left(m^2c^2 \Sigma^j -imc\beta
\epsilon^{jkl}\alpha_k\hat{p}_{l}+\Sigma^k\hat{p}_k\hat{p}^j
 \right)$ &
$\frac{\hbar}{2\hat p^0}\beta\left(mc\Sigma^j+\frac{\hat{p}^k\Sigma_k \hat p^j}{(\hat p^0+mc)}\right)$ &$s^j_{BMT}$
\\
\end{tabular*}
\end{center}
\end{center}
\end{table*}

To compare our operators with known in the literature, we remind that Pryce \cite{pryce1948mass} wrote his operators
acting on space of Dirac spinor $\Psi_D$, see the first column in Table \ref{tabular:Pryce-cm}. Foldy and Wouthuysen
\cite{foldy:1978} found unitary transformation which maps the Dirac equation
$i\hbar\partial_t\Psi_D=c(\alpha^ip_i+mc\beta)\Psi_D$ into the pair of square-root equations
$i\hbar\partial_t\Psi=c\beta\hat p^0\Psi$. Applying the FW transformation, the Pryce operators acquire block-diagonal
form on space $\Psi$, see the second column. Our operators act on space of solutions of square-root equation
(\ref{schrodinger-eq-bmt}), so we compare them with positive-energy parts (upper-left blocks) of Pryce operators of the
second column.

Our operators of canonical variables $\hat{\tilde{X}}^j=x^j$ and $\hat{\tilde{S}}^j$ correspond to the Pryce (e)
($\sim$ Foldy-Wouthuysen $\sim$ Newton-Wigner) position and spin operators.

However, operators of position $x^j$ and spin $S^j$ of our model are $\hat{X}^j$ and $\hat S^j$. They correspond to the
Pryce (d)-operators.

Operator of BMT-vector $\hat{S}_{BMT}^j$ is the Pryce (c) spin.

While we have started from relativistic theory (\ref{s.15}), working with the physical variables we have loosed, from the beginning, the
manifest relativistic covariance. Whether the quantum mechanics thus obtained is a relativistic theory? Below
we present a manifestly covariant formalism and confirm that scalar products, mean values and
transition probabilities can be computed in a covariant form.

\section{Minimal action in covariant formalism. Covariant form of noncommutative algebra of positions \label{sec:min-action-hamilton}}
Obtaining the minimal action (\ref{intr.5}) we have made various tricks. So, let us confirm that the action indeed leads to
the desired constraints (\ref{intr.2}) and (\ref{intr.3}). Computing conjugate momenta we obtain the primary constraint
$\pi_{g4}=0$, and the expressions
\begin{eqnarray}\label{ha.3}
p^\mu=mc\frac{N\dot x^\mu}{\sqrt{-\dot xN\dot x}}, \quad \pi^\mu=\sqrt{a_3}\frac{N\dot\omega^\mu}{\sqrt{\dot\omega
N\dot\omega}}. \nonumber
\end{eqnarray}
Due to Eq. (\ref{s.9}), they imply more primary constraints,
$p\omega=0$, $p^2+(mc)^2=0$, $\omega\pi=0$, and $\pi^2-a_3=0$.
Computing the Hamiltonian, $P\dot Q-L+\lambda_a\Phi_a$, we obtain
\begin{eqnarray}\label{ha.6}
& H&=\frac12\lambda_1(p^2+m^2c^2)+\lambda_3(\pi^2-a_3)+\frac{1}{2}g_4(\omega^2-a_4)+\cr
&\,& \lambda_5(\omega\pi)+\lambda_6(p\omega)+\lambda_4\pi_{g4}.
\end{eqnarray}
Preservation in time of the primary constraints implies the following chains of algebraic consequences:
\begin{eqnarray}\label{ha.7}
\pi_{g4}=0,  ~ \Rightarrow ~ \omega^2-a_4=0, ~ \Rightarrow ~ \lambda_5=0. \cr (\omega\pi)=0, ~ \Rightarrow ~
\lambda_3=\frac{a_4}{2a_3}g_4. \qquad \qquad \cr (p\omega)=0, ~ \Rightarrow ~ (p\pi)=0, ~ \Rightarrow ~ \lambda_6=0. \nonumber
\end{eqnarray}
As the result, the minimal action generates all the desired constraints (\ref{intr.2}) and (\ref{intr.3}). Three
Lagrangian multipliers have been determined in the process, $\lambda_5=\lambda_6=0$ and
$\lambda_3=\frac{a_4}{2a_3}g_4$, whereas $\lambda_1$ and $\lambda_4$ remain an arbitrary functions.

Besides the constraints, the action implies the Hamiltonian equations
$\dot g_4=\lambda_{4}$, $\dot\pi_{g4}=0$,
$\dot x^\mu=\lambda_1p^\mu$, $\dot p^\mu=0$,
$\dot\omega^\mu=\frac{a_4}{a_3}g_4\pi^\mu$, $\dot\pi^\mu=-g_4\omega^\mu$.
General solution to these equations in an
arbitrary and proper-time parameterizations is presented in Appendix 2.

To take into account the second-class constraints $T_4, T_5, T_6$ and $T_7$, we pass from Poisson to Dirac bracket. We
write them for the spin-plane invariant variables, they are $x^\mu$, $p^\mu$ and either the Frenkel spin-tensor or BMT
four-vector (\ref{ha.10.1}). The non vanishing Dirac brackets are as follows. \par \noindent Spacial sector:
\begin{equation}\label{db3}
\{x^\mu,x^\nu\}=-\frac{1}{2p^2}J^{\mu\nu}\,,\quad \{x^\mu,p^\nu\}=\eta^{\mu\nu}, \quad \{p^\mu,p^\nu\}=0.
\end{equation}
\par\noindent Frenkel sector:
\begin{equation}\label{db5}
\{J^{\mu\nu},J^{\alpha\beta}\}= 2(g^{\mu\alpha} J^{\nu\beta}-g^{\mu\beta} J^{\nu\alpha}-g^{\nu\alpha} J^{\mu\beta}
+g^{\nu\beta} J^{\mu\alpha})\,,
\end{equation}
\begin{equation}\label{db6}
\{x^\mu,J^{\alpha\beta}\}=\frac{1}{p^2}J^{\mu[\alpha}p^{\beta]} \,,
\end{equation}
\par \noindent BMT-sector:
\begin{eqnarray}\label{db7}
\{s^\mu, s^\nu\}=-\frac{1}{\sqrt{-p^2}}\epsilon^{\mu\nu\alpha\beta}p_\alpha s_\beta=\frac{1}{2}J^{\mu\nu},
\end{eqnarray}
\begin{eqnarray}\label{db8}
\{x^\mu, s^\nu\}=-\frac{s^\mu p^\nu}{p^2}=-\frac{1}{4\sqrt{-p^2}}\epsilon^{\mu\nu\alpha\beta}J_{\alpha\beta}-\frac{p^\mu
s^\nu}{p^2}.
\end{eqnarray}
In the equation (\ref{db5}) it has been denoted
$g^\mu{}_\nu\equiv\delta^\mu{}_\nu-\frac{p^{\mu}p_\nu}{p^2}$.
Together with $\tilde g^\mu{}_\nu\equiv\frac{p^\mu p_\nu}{p^2}$, this forms a pair of projectors
$g+\tilde g=1$, $g^2=g$, $\tilde g^2=\tilde g$, $g\tilde g=0$.
The transition to spin-plane invariant variables does not spoil manifest covariance. So, we write equations
of motion in terms of these variables
\begin{eqnarray}\label{ha.9}
\dot x^\mu=\lambda_1p^\mu, \qquad \dot p^\mu>0,
\end{eqnarray}
\begin{eqnarray}\label{ha.19.1}
\dot J^{\mu\nu}=0, \quad J^{\mu\nu}p_\nu=0, \quad J^2=6\hbar^2.
\end{eqnarray}
\begin{eqnarray}\label{ha.20.1}
\dot S^\mu=0, \quad S^\mu p_\nu=0, \quad S^2=\frac{3\hbar^2}{4}.
\end{eqnarray}
Besides, we have the first-class constraint
\begin{eqnarray}\label{ha.20.2}
p^2+(mc)^2=0\,, \quad \mbox{where} \quad p^0>0.
\end{eqnarray}

Let us compare these results with non manifestly covariant formalism of previous section. Evolution of physical
variables can be obtained from equations (\ref{ha.9})-(\ref{ha.20.2}) assuming that the functions $Q^\mu(\tau)$
represent the physical variables $Q^i(t)$ in the parametric form. Using the formula $\frac{dF}{dt}=c\frac{\dot
F(\tau)}{\dot x^0(\tau)}$, this gives  Eqs. (\ref{pha.8}), (\ref{pha.12}) and (\ref{pha.14}). The brackets
(\ref{pha.15})-(\ref{pha.17}) of physical variables appeared, if we impose the physical-time gauge $x^0-\tau=0$ for the
constraint (\ref{ha.20.2}), and pass from (\ref{db3})-(\ref{db8}) to the Dirac bracket which take into account this
second-class pair. Physical Hamiltonian (\ref{pha.19}) can be obtained from (\ref{ha.6}) considering the physical-time
gauge as a canonical transformation \cite{gitman1990quantization}.

Summarizing, in classical mechanics all basic relations for physical variables can be obtained from covariant
formalism. In the next section we discuss, how far we can proceed towards formulation of quantum mechanics in a
manifestly-covariant form.

\section{Manifestly-covariant form of quantum mechanics of the Frenkel electron\label{sec:RQM}}
According to Wigner \cite{wigner1939unitary, Bargmann1948, Weinberg_2009}, with an elementary particle in QFT we
associate the Hilbert space of representation of Poincare group. The space can be described in a manifestly covariant
form as a space of solutions to Klein-Gordon (KG) equation for properly chosen multicomponent field $\psi_i(x^\mu)$.
One-component field corresponds to spin-zero particle. Two-component field has been considered by Feynman and Gell-Mann
\cite{feynman1958fermi-interaction} to describe weak interaction of spin one-half particle, and by Brown as a starting
point for QED \cite{brown1958two-component-fermi}. It is well-known, that one-component KG field has no
quantum-mechanical interpretation. In contrast, two-component KG equation does admit the probabilistic interpretation:
the four-vector (\ref{cq17}) represents positively defined conserved current of this equation. On this base, we
consider below the relativistic quantum mechanics of two-component KG equation and show its equivalence with quantum
mechanics of Dirac equation. Then we show that the covariantly quantized F-BMT electron corresponds to positive-energy
sector of this quantum mechanics. At last, we establish the correspondence between canonical and covariant
formulations, thus proving relativistic invariance of the physical-time formalism of subsection
\ref{sec:min-action-physical_2}.

\subsection{Relativistic quantum mechanics of two-component Klein-Gordon equation \label{subsec:KG}}
We denote states and operators of covariant formalism by small letters, to distinguish them from the quantities of
canonical formalism. Consider the space of abstract state-vectors composed by two-component Weyl spinors
$\psi_a(x^\mu)$, $a=1, 2$. Generators of Poincare transformations in this space read
\begin{eqnarray}\label{cq9.1}
\hat{m}^{\mu\nu}=x^\mu\hat p^\nu-x^\nu\hat p^\mu+ \frac{1}{2}\sigma^{\mu\nu}\,, \qquad \hat p_\mu=-i\hbar\partial_\mu\,,
\end{eqnarray}
where the Lorentz generators
\begin{eqnarray}\label{cq10}
\sigma^{\mu\nu}=-\frac{i\hbar}{2}(\sigma^\mu\bar\sigma^\nu-\sigma^\nu\bar\sigma^\mu), \nonumber
\end{eqnarray}
are built from standard Pauli matrices $\sigma^i$ combined into the sets
\begin{eqnarray}\label{cq8}
\sigma^\mu=({\bf 1}, \sigma^i), \qquad \bar\sigma^\mu=(-{\bf 1}, \sigma^i). \nonumber
\end{eqnarray}
They are hermitian and obey
$\sigma^\mu\bar\sigma^\nu+\sigma^\nu\bar\sigma^\mu=2\eta^{\mu\nu}$,
$\bar\sigma^\mu\sigma^\nu+\bar\sigma^\nu\sigma^\mu=2\eta^{\mu\nu}$.
Further, on the Poincare-invariant subspace selected by two-component KG equation
\begin{eqnarray}\label{cq16}
(\hat p^2+m^2c^2)\psi=0\,,
\end{eqnarray}
we define an invariant and positive-defined scalar product as follows. The four-vector\footnote{$\dagger$ denotes usual
Hermitian conjugation, $\hat a^\dagger=(\hat a^*)^T$, $(\hat a\hat b)^\dagger=\hat b^\dagger \hat a^\dagger$, then
$(\hat p^\mu f)^\dagger=-\hat p^\mu (f)^\dagger$.}
\begin{eqnarray}\label{cq17}
I^\mu[\psi,\phi]=\frac{1}{m^2c^2}(\bar\sigma\hat p\psi)^\dagger\sigma^\mu\bar\sigma\hat p\phi- \psi^\dagger\bar\sigma^\mu\phi\,,
\end{eqnarray}
represents a conserved current of Eq. (\ref{cq16}), that is $\partial_\mu I^{\mu}=0$, when $\psi$ and $\phi$ satisfy to
Eq. (\ref{cq16}).  Then the integral
\begin{equation}\label{cq-inv-scalar-product}
(\psi,\phi)=\int\limits_{\Omega} d\Omega_\mu I^\mu\,,\qquad d\Omega_\mu=\frac{d^4x}{dx_\mu}\,,
\end{equation}
does not depend on the choice of a space-like 3-dimensional hyperplane $\Omega$ (an inertial coordinate
system). As a consequence, this does not depend on time. So we can restrict ourselves to the hyperplane $\Omega$ defined by
the equation $x^0=\mbox{const}$, then
\begin{equation}\label{sp1}
(\psi,\phi)=\int d^3x I^0\,.
\end{equation}
Besides, this scalar product is positive-defined\footnote{See also a detailed discussion of positively defined scalar products for the Klein-Gordon-type equations \cite{mostafazadeh2003hilbert}}, since
\begin{equation}\label{sp2}
I^0[\psi,\psi]=\frac{1}{m^2c^2}(\bar\sigma\hat p\psi)^\dagger\bar\sigma\hat p\psi+\psi^\dagger \psi >0.
\end{equation}
So, this can be considered as a probability density of operator $\hat{\bf x}={\bf x}$. We point out that transformation
properties of the column $\psi$ are in the agreement with this scalar product: if $\psi$ transforms as a (right) Weyl
spinor, then $I^\mu$ represents a four-vector.

Now we can confirm relativistic invariance of scalar product (\ref{fw-scalar-product}) of canonical formalism. The operator
$\hat p^0$ is hermitian on the subspace of positive-energy solutions $\psi$, so we can write
\begin{eqnarray}\label{sp3}
(\psi, \phi)=\int d^3x \frac{1}{m^2c^2}(\bar\sigma\hat p\psi)^\dagger\bar\sigma\hat p\phi+\psi^\dagger\bar\phi =\\
\int
d^3x \left[\left(\frac{1}{mc}\bar\sigma\hat p+i\right)\psi\right]^\dagger\left(\frac{1}{mc}\bar\sigma\hat
p+i\right)\phi\,, \nonumber
\end{eqnarray}
This suggests the map $W: \{\psi\}  \rightarrow  \{\Psi\}$, $\Psi=W\psi$,
\begin{eqnarray}\label{sp4}
W=\frac{\bar\sigma\hat p}{mc}+i\,, \quad W^{-1}=\frac{1}{2\hat p^0}\left(i\sigma\hat
p-mc\right)\,,
\end{eqnarray}
which respects the scalar products (\ref{fw-scalar-product}) and (\ref{sp1}), and thus proves relativistic invariance
of the scalar product $\langle \Psi,\Phi \rangle$
\begin{eqnarray}\label{sp5}
\langle \Psi,\Phi \rangle=(\psi, \phi)\,.
\end{eqnarray}
We note that map $W$ is determined up to an isometry, we can multiply $W$ from the left by an arbitrary unitary operator $U$,
$W\to W'=U W$, $U^\dagger U=1$. Here $\dagger$ denotes Hermitian conjugation with respect to scalar product $\langle,\rangle$.
The ambiguity in the definition of $W$ can be removed by the polar decomposition of the operator \cite{conway1990course}.
A bounded operator between Hilbert spaces admits the following factorization: $W=PV$, where $V=(W^\dagger W)^{1/2}$, $P=WV^{-1}$.
Positively defined operator $W^\dagger W>0$ has a unique square root
$(W^\dagger W)^{1/2}$.
Moreover $W^\dagger W=W'^\dagger W'$, therefore $V$ defines map from $\{\psi\}$ to $\{\Psi\}$ without ambiguity.
We present the explicit form of $V$ in subsection \ref{covinv}.

\subsection{Relation with Dirac equation \label{subsec:QM-BMT-Dirac-equiv}}
Here we demonstrate equivalence of quantum mechanics of KG and Dirac equations. To this aim, let us replace two
equations of second order, (\ref{cq16}), by equivalent system of four equations of the first order. To achieve this,
with the aid of the identity $\hat p^\mu \hat p_\mu=\sigma^\mu \hat p_\mu\bar\sigma^\nu\hat p_\nu$, we represent
(\ref{cq16}) in the form
\begin{eqnarray}\label{cq18}
\sigma^\mu \hat p_\mu\bar\sigma^\nu\hat p_\nu \psi+m^2c^2\psi=0.
\end{eqnarray}
Consider an auxiliary two-component function $\bar\xi$ (Weyl spinor of opposite chirality), and define evolution of
$\psi$ and $\bar\xi$ according the equations \footnote{Note that
$\bar\xi$ can be considered as conjugated momentum for $\psi$, than the passage from (\ref{cq18}) to (\ref{cq20}) is
just the passage from a Lagrangian to Hamiltonian formulation. Similar interpretation can be developed for the
Schrodinger equation, see \cite{AAD.Schrodinger}.}
\begin{eqnarray}\label{cq19}
\sigma^\mu \hat p_\mu(\bar\sigma^\nu\hat p_\nu) \psi+m^2c^2\psi=0, \\
(\bar\sigma^\nu\hat p_\nu)\psi-mc\bar\xi=0.\label{cq19-2}
\end{eqnarray}
That is dynamics of $\psi$ is determined by (\ref{cq18}), while $\bar\xi$ accompanies $\psi$: $\bar\xi$ is determined
from the known $\psi$ taking its derivative, $\bar\xi=\frac{1}{mc}(\bar\sigma\hat p)\psi$. Evidently, the systems
(\ref{cq16}) and (\ref{cq19}), (\ref{cq19-2}) are equivalent.  Rewriting the system (\ref{cq19}), (\ref{cq19-2}) in a
more symmetric form, we recognize the Dirac equation
\begin{eqnarray}\label{cq20}
\left(
\begin{array}{cc}
0& \sigma^\mu\hat p_\mu\\
-\bar\sigma^\nu\hat p_\nu & 0
\end{array}
\right)\left(
\begin{array}{c}
\psi\\
\bar\xi
\end{array}
\right)+mc\left(
\begin{array}{c}
\psi\\
\bar\xi
\end{array}
\right)=0,\\ 
(\gamma_W^\mu\hat p_\mu+mc)\Psi=0\,,\nonumber
\end{eqnarray}
for the Dirac spinor $\Psi=\left(\psi, \bar\xi\right)$
in the Weyl representation of $\gamma$\,-matrices
\begin{eqnarray}\label{cq22}
\gamma_W^0=\left(
\begin{array}{cc}
0 & {\bf 1}\\
{\bf 1} & 0
\end{array}
\right), \qquad \gamma_W^i=\left(
\begin{array}{cc}
0& \sigma^i\\
-\sigma^i& 0
\end{array}
\right). \nonumber
\end{eqnarray}
This gives one-to-one correspondence among two spaces. With each solution $\psi$ to KG equation we associate the solution
\begin{eqnarray}\label{cq21}
\Psi[\psi]=\left(
\begin{array}{c}
\psi\\
\frac{1}{mc}(\bar\sigma\hat p)\psi
\end{array}
\right), \nonumber
\end{eqnarray}
to the Dirac equation. Below we also use the Dirac representation of $\gamma$\,-matrices
\begin{eqnarray}\label{cq22}
\gamma^0=\left(
\begin{array}{cc}
{\bf 1} & 0\\
0&    -{\bf 1}
\end{array}
\right), \qquad \gamma^i=\left(
\begin{array}{cc}
0& \sigma^i\\
- \sigma^i& 0
\end{array}
\right).
\end{eqnarray}
In this representation, the Dirac spinor corresponding to $\psi$ reads
\begin{eqnarray}\label{cq22.1}
\Psi_D[\psi]=\frac{1}{\sqrt{2}}\left(
\begin{array}{cc}
\,\,\,\,{\bf 1} & {\bf 1}\\
-{\bf 1} & {\bf 1}
\end{array}
\right)\left(
\begin{array}{c}
\psi\\
\frac{1}{mc}(\bar\sigma\hat p)\psi
\end{array}
\right)=\\
\frac{1}{\sqrt{2}mc}\left(
\begin{array}{c}
{[(\bar\sigma\hat p)+mc]}\psi\\
{[(\bar\sigma\hat p)-mc]}\psi
\end{array}
\right).\nonumber
\end{eqnarray}
The conserved current (\ref{cq17}) of KG equation (\ref{cq16}), being rewritten in terms of Dirac spinor, coincides with the Dirac current
\begin{eqnarray}\label{cq23}
I^\mu[\psi_1,\psi_2]=\bar\Psi[\psi_1]\gamma^\mu\Psi[\psi_2].
\end{eqnarray}
Therefore, the scalar product (\ref{cq-inv-scalar-product}) coincides with that of Dirac.

\subsection{Covariant operators of F-BMT electron\label{covop}}
In a covariant scheme, we need to construct operators $\hat{x}^\mu, \hat p^\mu, \hat j^{\mu\nu}, \hat s_{BMT}^\mu$ whose
commutators
\begin{eqnarray}\label{cq23.1}
[\hat q_1 , \hat q_2 ]=i\hbar \left.\{ q_1 , q_2 \}_D\right|_{q_i\rightarrow\hat q_i},
\end{eqnarray}
are defined by the Dirac brackets (\ref{db3})-(\ref{db8}). Inspection of the classical equations
$S^2=\frac{3\hbar^2}{4}$ and $p^2+(mc)^2=0$ suggests that we can look for a realization of operators in the Hilbert space
constructed in subsection \ref{subsec:KG}.

With the spin-sector variables we associate the operators
\begin{equation}\label{cq-S-2operator}
s_{BMT}^\mu \quad \rightarrow \quad \hat s_{BMT}^\mu=\frac{1}{4\sqrt{-\hat p^2}}\epsilon^{\mu\nu\alpha\beta}\hat p_\nu
\sigma_{\alpha\beta} \,,
\end{equation}
\begin{eqnarray}\label{cq11}
J^{\mu\nu} \quad \rightarrow \quad \hat j^{\mu\nu}\equiv -\frac{2}{\sqrt{-\hat p^2}}\epsilon^{\mu\nu\alpha\beta}\hat
p_\alpha \hat s_{BMT \beta}=\\
\sigma^{\mu\nu}+\frac{\hat p^\mu(\sigma \hat p)^\nu-\hat p^\nu(\sigma \hat p)^\mu}{\hat
p^2}\,.\nonumber
\end{eqnarray}
%
%
They obey the desired commutators (\ref{cq23.1}), (\ref{db7}), (\ref{db5}). To find the position operator, we separate
the inner angular momentum $\hat j^{\mu\nu}$ in the expression (\ref{cq9.1}) of Poincare generator
\begin{eqnarray}\label{tl-part-of-hatM}
\hat m^{\mu\nu} =\left[x^\mu+\frac{(\sigma \hat p)^\mu }{2\hat p^2}\right]\! \hat p^\nu\!-\!\left[x^\nu+\frac{(\sigma \hat
p)^\nu }{2p^2}\right]\!\hat p^\mu\!+\frac12\hat j^{\mu\nu}\,.
\end{eqnarray}
This suggests the operator of ``relativistic position''\footnote{Classical analog of this operator also appeared as a
gauge-invariant variable in mechanical model of Dirac equation, see \cite{AAD5}.}
\begin{eqnarray}\label{cq13}
x^\mu \quad \rightarrow \quad \hat x_{rp}^\mu=\hat x^\mu+\frac{1}{2\hat p^2}(\sigma \hat p)^\mu\,,
\end{eqnarray}
where $\hat x^\mu\psi=x^\mu\psi$. The operators $\hat p_\mu=-i\hbar\partial_\mu$, (\ref{cq-S-2operator}),
(\ref{cq11}) and (\ref{cq13}) obey the algebra (\ref{cq23.1}),
(\ref{db3})-(\ref{db8}).

Equation (\ref{ha.20.1}) in this realization states that square of second Casimir of Poincare group has fixed value
$\frac{3\hbar^2}{4}$, and in the representation chosen is satisfied identically. The equations
(\ref{ha.20.2}) just state that we work in the positive-energy subspace of the Hilbert space of KG equation (\ref{cq16}).

We thus completed our covariant quantization procedure by matching classical variables of reparametrization-invariant
formulation to operators acting on the Hilbert space of two component spinors with scalar product
(\ref{cq-inv-scalar-product}). The construction presented is manifestly Poincare-covariant. In the next subsection we
discuss the connection between canonical and manifestly covariant formulations of the F-BMT electron.

\subsection{Relativistic invariance of canonical formalism\label{covinv}}
Relativistic invariance of the scalar product (\ref{fw-scalar-product}) has been already shown in subsection
\ref{subsec:KG}. Here we show how the covariant formalism can be used to compute mean values and probability rates of
canonical formulation, thus proving its relativistic covariance. Namely, we confirm the following \par \noindent

{\bf Proposition.} Let
\begin{eqnarray}\label{cor0}
H^+_{can} &=& \left\lbrace ~\Psi(t, \vec x)\,; ~  i\hbar\frac{d\Psi}{dt}=\sqrt{\hat{{\bf p}}^{\,2}+(mc)^2}\Psi\,,\right.\\ &\, & \left. \langle \Psi,\Phi
\rangle=\int d^3x\Psi^\dagger \Phi ~ \right\rbrace \,, \nonumber
\end{eqnarray}
is Hilbert space of canonical formulation and
\begin{eqnarray}\label{cor0.1}
H_{cov} &=&\left\lbrace ~ \psi(x^\mu)\,; ~ (\hat p^2+m^2c^2)\psi=0\,, \right. \\ 
&\, & \left.  (\psi,\phi)=\int\limits_{\Omega} d\Omega_\mu I^\mu[\psi,\phi]\,, ~
\right\rbrace \,, \nonumber
\end{eqnarray}
is Hilbert space of two-component KG equation. 

With a state-vector $\Psi$ we associate $\psi$ as follows:
\begin{eqnarray}\label{cor0.2}
\psi=V^{-1}\Psi\,, \quad V^{-1}=\frac{1}{2\sqrt{\hat p^0(\hat p^0+mc)}}[mc-\sigma\hat p]\,.
\end{eqnarray}
Then $\langle \Psi,\Phi \rangle=(\psi, \phi )$. Besides, mean values of the physical position and spin operators
(\ref{electron-position-d-components-FW})-(\ref{spin.1}) can be computed as follows
\begin{eqnarray}\label{cor0.3}
\langle \Psi,\hat X^i\Phi \rangle=\mbox{Re}(\psi, \hat x_{rp}^i\phi ), \qquad \langle \Psi,\hat J^{ij}\Phi
\rangle=(\psi, \hat j^{ij}\phi )\,, \cr 
\langle \Psi,\hat S^i\Phi \rangle=\frac14\epsilon^{ijk}(\psi, \hat
j^{jk}\phi )\,, \nonumber
\end{eqnarray}
where  $\hat x_{rp}^i$ and $\hat j^{ij}$ are spacial components of the manifestly-covariant operators
\begin{eqnarray}\label{cor0.4}
\hat x_{rp}^\mu=\hat x^\mu+\frac{(\sigma \hat p)^\mu}{2\hat p^2}\,, \quad \hat j^{\mu\nu}=\sigma^{\mu\nu}+\frac{\hat
p^\mu(\sigma \hat p)^\nu-\hat p^\nu(\sigma \hat p)^\mu}{\hat p^2}\,. \nonumber
\end{eqnarray}
We also show that the map $V$ can be identified with Foldy-Wouthuysen transformation applied to the Dirac spinor
(\ref{cq22.1}).

It will be convenient to work in the momentum representation, $\psi(x^\mu)=\int d^4p \psi(p^\mu){\rm e}^{\frac{i}{\hbar}px}$.
Transition to the momentum representation implies the substitution
\[
\hat p_\mu \to p_\mu\,,\qquad \hat x_\mu \to i\hbar \frac{\partial}{\partial p^\mu}\,,
\]
in the expressions of covariant operators (\ref{cq-S-2operator}), (\ref{cq11}), (\ref{cq13}) and so on.

An arbitrary solution to the KG equation reads
\[
\psi(t, {\bf x})=\int d^3p \left(\psi({\bf p}){\rm e}^{\frac{i\omega_px^0}{\hbar}}+\psi_{-}({\bf p}){\rm e}^{\frac{-i\omega_px^0}{\hbar}}\right){\rm e}^{-\frac{i({\bf p x})}{\hbar}}\,,
\]
\[ \omega_p\equiv\sqrt{{\bf p}^2+(mc)^2}\,,
\]
where $\psi({\bf p})$ and $\psi_{-}({\bf p})$ are arbitrary functions of three-momentum, they correspond to positive
and negative energy solutions. The scalar product can
be written then as follows
\[
(\psi,\phi)=2\int \frac{d^3p\, \omega_p}{m^2c^2} \left[\psi^\dagger(\bar\sigma p)\phi-\psi_{-}^\dagger(\sigma
p)\phi_{-}\right] \,, 
\]
where
\[ 
(\bar\sigma p)=\omega_p+({\boldsymbol{\sigma} \bf p})\,,\quad (\sigma p)=-\omega_p+({\boldsymbol{\sigma} \bf p})\,.
\]
We see that this scalar product separates positive and negative energy parts of state vectors. Since our classical
theory contains only positive energies, we restrict our further considerations by the positive energy solutions only.
In the result, in the momentum representation the scalar product (\ref{sp1}) reads in terms of non-trivial metric
$\rho$ as follows:
\begin{equation}\label{cq-hilber-space-matric}
(\psi,\phi)=\int d^3p\psi^\dagger\rho\phi, \qquad \rho=\frac{2\omega_p}{m^2c^2}(\bar\sigma p)\,.
\end{equation}
Now our basic space is composed by arbitrary functions $\psi({\bf p})$. The operators $\hat x^i$, $\hat s^\mu$ and
$\hat j^{\mu\nu}$ act on this space as before, with the only modification, that $\hat p^0\psi({\bf
p})=\omega_p\psi({\bf p})$. The operator $\hat x^0$ and, as a consequence, the operator $\hat x_{rp}^0$, do not act in
this space. Fortunately, they are not necessary to prove the proposition formulated above.

Given operator $\hat A$ we denote its hermitian conjugated in space $H^+_{can}$ as $\hat A^\dagger$. Hermitian
operators in space $H^+_{can}$ have both real eigenvalues and expectation values.
%
Consider an operator $\hat a$ in space $H_{cov}$
with  real expectation values
$(\psi,\hat a\psi)=(\psi,\hat a\psi)^*$.
It should obey
$\hat a^\dagger \rho =\rho \hat a$.
That is, such an operator in $H_{cov}$ should be pseudo-Hermitian. We denote pseudo-Hermitian conjugation in $H_{cov}$
as follows:
$\hat a_c=\rho^{-1}\hat a^\dagger \rho$.
Then pseudo-Hermitian part of an operator $\hat a$ is given by
$\frac12(\hat a+\hat a_c)$.

Let us check the pseudo-Hermicity properties of basic operators. From the following identities:
\[
(\sigma^{\mu\nu})^\dagger\rho=\rho\,\left(\sigma^{\mu\nu}+\frac{2i\hbar}{p^2} (\sigma p) (p^\mu \bar\sigma^\nu- p^\nu
\bar\sigma^\mu)\right)\,, \quad
\]
\[
(\sigma^{\mu\nu} p_\nu)^\dagger\rho=\rho\,\left(\sigma^{\mu\nu} p_\nu+2i\hbar [ p^\mu - (\sigma p)
\bar\sigma^\mu]\right)\,,
\]
\[
(\hat x_{rp}^j)^\dagger\rho=\rho \left(\hat x_{rp}^j+\frac{i\hbar}{m^2c^2\omega_p}
\left[\frac{m^2c^2}{\omega_p}p^j-p^j(\vec{\sigma}\vec{p})\right] \right)\,,
\]
we see that operators $\sigma^{\mu\nu}$ and $\hat x^j_{rp}$ are non-pseudo-Hermitian, while operators $\hat p^\mu$, $\hat
s^\mu$, $\hat j^{\mu\nu}$ and orbital part of $\hat m^{ij}$ are pseudo-Hermitian.

To construct the map (\ref{cor0.2}) we look for square root of the metric, $V=\rho^{1/2}$. Metric $\rho$ is positively
defined, therefore the square root is unique \cite{conway1990course}, this reads
\begin{equation}\label{opV}
V=\frac{1}{mc}\sqrt{\frac{\omega_p}{\omega_p+mc}}[(\bar \sigma p)+mc]\,.
\end{equation}
We use this to define the map $H_{cov} \to H^+_{can}$, ~ $\Psi=V\psi$, which corresponds to
the polar decomposition of map $W$ defined in \eqref{sp4}.
Then the scalar product (\ref{cq-hilber-space-matric}) can be rewritten as
\[
(\psi, \phi)=\int d^3p(V\psi)^\dagger V\phi=\int d^3p\Psi^\dagger\Phi=\langle \Psi,\Phi
\rangle\,.
\]
This proves relativistic invariance of the scalar product $\langle \Psi,\Phi \rangle$ of canonical formalism.

Our map defined by operator $V$ turns out to be in the close relation with the Foldy-Wouthuysen transformation.
It can be seen applying the Foldy-Wouthuysen unitary transformation
\[
U_{FW}=\frac{\omega_p+mc+(\vec{\gamma}\vec{p})}{\sqrt{2(\omega_p+mc)\omega_p}},
\]
to the  Dirac spinor $\Psi_D[\psi]$,
\[
\Psi_{FW}[\psi]=U_{FW}\Psi_D[\psi]=
\left(
\begin{array}{c}
V\psi\\
0
\end{array}
\right)=\left(
\begin{array}{c}
\Psi\\
0
\end{array}
\right)\,.
\]
The last equation means that operator $V$ is a restriction of operator $U_{FW}$
to the space of positive-energy right Weyl spinors $\psi$.

The transformation between state-vectors induces the map of operators
\begin{equation}\label{cor4}
\hat Q=V\hat qV^{-1}\,,
\end{equation}
where
\[ V^{-1}=\frac{1}{2\sqrt{\omega_p(\omega_p+mc)}}[mc-(\sigma p)]\,.\] Then
\begin{equation}\label{cor5}
\langle \Psi,\hat Q\Phi \rangle=(\psi, \hat q\phi)\,.
\end{equation}
Due to Hermicity of $V$, $V^\dagger=V$, pseudo-Hermitian operators,
$\hat q^\dagger V^2= V^2 \hat q$,
transform into Hermitian operators
$\hat Q^\dagger = \hat Q$.
For an operator $\hat q$ which commutes with momentum operator, transformation \eqref{cor4}
acquire the following form
\[
\hat Q=\frac{1}{2}(\hat q+\hat q^\dagger)-\frac{1}{2(\omega_p+mc)}(\hat q-\hat q^\dagger)(\vec{\sigma}\vec{p})\,.
\]
Using this formula, we have checked by direct computations that covariant operators $\hat{\bf{p}}$, $\hat j^{\mu\nu}$
and $\hat s_{BMT}^\mu$ transform into canonical operators $\hat{\bf p}$, $\hat J^{\mu\nu}$ and $\hat S_{BMT}^\mu$, so
the spacial part of $\hat J^{\mu\nu}$, $\hat S^i=\frac14\epsilon^{ijk}\hat J_{jk}$ represents the classical spin $S^i$.
This observation together with Eq. (\ref{cor5}) implies that mean values of the operators of canonical formalism are
relativistic-covariant quantities.

Concerning the position operator, we first apply the inverse to Eq. \eqref{cor4} to our canonical
coordinate $\hat{\tilde X}^i=i\hbar \frac{\partial}{\partial p^i}$ in the momentum representation
\[
\hat{\tilde x}_V^i=V^{-1}\hat{\tilde X}^iV=\hat{\tilde X}^i+[V^{-1},\hat{\tilde X}^i]V=
\]
\[
i\hbar \frac{\partial}{\partial p^i}-\frac{i\hbar p^i(\vec{\sigma}\vec{p})}{2mc\omega_p(\omega_p+mc)}
+\frac{i\hbar p^i}{2\omega_p}+
\frac{i\hbar\sigma^i}{2mc}
+\frac{\hbar\epsilon^{ijk}\sigma_jp_k}{2mc(\omega_p+mc)}\,.
\]
Our position operator then can be mapped as follows:
\begin{eqnarray}
\hat{x}_V^i= V^{-1}\left(i\hbar \frac{\partial}{\partial p^i}+\frac{\epsilon^{ijk}\hat
S_jP_k}{mc(\omega_p+mc)}\right)V =\nonumber\\
\label{cov-position-d}
i\hbar \frac{\partial}{\partial p^i}+\frac{i\hbar p^i(\vec{\sigma}\vec{p})}{2p^2\omega_p} +\frac{i\hbar
p^i}{2\omega_p}- \frac{i\hbar\omega_p\sigma^i}{2p^2} +\frac{\hbar\epsilon^{ijk}p_j\sigma_k}{2p^2}\,. 
\end{eqnarray}
We note that pseudo-Hermitian part of operator $\hat x_{rp}^i$
coincides with the image $\hat{x}_V^i$,
\[
\hat{x}_V^i=\frac{1}{2}\left(\hat x_{rp}^i+\left[\hat x_{rp}^i\right]_c\right)\,.
\]
Since $\hat x_{rp}^\mu$ has explicitly covariant form, this also proves covariant
character of position operator $\hat{X}^i$. Indeed, \eqref{cor4} means that
matrix elements of $\hat{X}^i$ are expressed through the real part of manifestly covariant matrix elements
\[
\langle \Psi,\hat{X}^i \Phi \rangle=(\psi, \hat x_V^i \phi)={\rm Re}(\psi, \hat{x}_{rp}^i \phi)\,.
\]
%
%
%
In summary, we have proved the proposition formulated above. The operators $\hat j^{\mu\nu}$ and $\hat{x}_{rp}^\mu$,
which act on the space of two-component KG equation, represent manifestly-covariant form of the Pryce (d)-operators.

Table \ref{tabular:non-covariantFW-covariant-quantization} summarizes manifest form of operators of canonical formalism
and their images in covariant formalism.
\begin{center}
\begin{table*}
\begin{center}
\caption{Operators of canonical and manifestly covariant formulations in momentum representation}
\label{tabular:non-covariantFW-covariant-quantization}
\begin{tabular*}{\textwidth}{c|c|c}
{}  & Canonical formalism $\Psi({\bf p})$         & Covariant formalism $\psi({\bf p})$\\
\hline \hline
$\hat P_j \to \hat p_j $   & $p_j$            & $p_j$
\\
$\hat S^i \to \hat s^i $   & $\frac{\hbar}{2mc}\left(\omega_p
\sigma^i-\frac{1}{(\omega_p+mc)}(\vec{p}\vec{\sigma})p^i\right)$            &
$\frac{\hbar\omega_p}{2(mc)^2}\left(\omega_p \sigma^i-(\vec{p}\vec{\sigma})p^i-i\epsilon_{imn}p^m\sigma^n\right)$
\\
$\hat X^i \to \hat x^i_V $  & $i\hbar \frac{\partial}{\partial
p^i}-\frac{\hbar}{2mc(\omega_p+mc)}\epsilon^{ijk}p_j\sigma_k$  &    $i\hbar \frac{\partial}{\partial p^i}+\frac{i\hbar
p^i(\vec{\sigma}\vec{p})}{2p^2\omega_p} +\frac{i\hbar p^i}{2\omega_p}- \frac{i\hbar}{2p^2}\omega_p\sigma^i
+\frac{\hbar}{2p^2}\epsilon^{ijk}p_j\sigma_k$
\\
$\hat J^{ij}\to \hat j^{ij} $& $\frac{\hbar}{mc}\epsilon^{ijk}\left(\omega_p
\sigma_k-\frac{1}{(\omega_p+mc)}(\vec{p}\vec{\sigma})p_k\right)$  &
$\frac{\hbar\omega_p}{m^2c^2}\epsilon^{ijk}\left(\omega_p
\sigma_k-(\vec{p}\vec{\sigma})p_k-i\epsilon_{kmn}p^m\sigma^n\right)$
\\
$\hat J^{0i}\to \hat j^{0i} $& $-\frac{\hbar}{mc}\epsilon^{ijk}p_j \sigma_k$  & $-\frac{\hbar}{m^2c^2}\epsilon^{ijk}
\left(\omega_p \sigma_k- i \epsilon_{kml}p^m\sigma^l\right)p_j$
\\
$\hat S_{BMT}^0 \to \hat s_{BMT}^0 $   & $\frac{\hbar}{2mc}(\vec{p}\vec{\sigma})$            &
$\frac{\hbar}{2mc}(\vec{p}\vec{\sigma})$
\\
$\hat S_{BMT}^i \to \hat s_{BMT}^i$   &
$\frac{\hbar}{2}\left(\sigma^i+\frac{1}{mc(\omega_p+mc)}(\vec{p}\vec{\sigma})p^i\right)$ &
$\frac{\hbar}{2mc}(\omega_p\sigma^i +i\epsilon^{ijk}p_j\sigma_k )$
\\
\end{tabular*}
\end{center}
\end{table*}
\end{center}

\subsection{Manifestly-covariant operators of spin and position of Dirac equation\label{covdir}}
According to Eq. (\ref{cq23}), the scalar product $(\psi, \phi)$ coincides with that of Dirac. This allows us to find
manifestly-covariant operators in the Dirac theory which have the same expectation values as $\hat{j}^{\mu\nu}$ and
$\hat{x}_{rp}^\mu$ . Consider the following analog of $\hat j^{\mu\nu}$ on the space of 4-component Dirac spinors
\begin{eqnarray}\label{J-hilderwood}
\hat{j}_D^{\mu\nu}=\Sigma^{\mu\nu}+\frac{\hat p^\mu\Sigma^{\nu\alpha} \hat p_\alpha-\hat p^\nu\Sigma^{\mu\alpha} \hat
p_\alpha}{\hat p^2}=\cr
\Sigma^{\mu\nu}+\frac{i\hbar}{\hat p^2}\left(\hat p^\mu\gamma^{\nu} -\hat
p^\nu\gamma^{\mu}\right)(\gamma\hat p) \,,
\end{eqnarray}
where $\Sigma^{\mu\nu}=\frac{i\hbar}{2}(\gamma^\mu\gamma^\nu-\gamma^\nu\gamma^\mu)$. This definition is independent
from a particular representation of $\gamma$-matrices. In the representation \eqref{cq22} this reads
\[
\Sigma^{\mu\nu}=\left(
\begin{array}{cc}
\sigma^{\mu\nu} & 0\\
0 & (\sigma^{\mu\nu})^\dagger
\end{array}
\right)\,,
\]
and can be used to prove the equality of matrix elements
\[
\int d^3x \Psi[\psi]^\dagger\hat{j}_D^{\mu\nu}\Phi[\phi]=(\psi, \hat{j}^{\mu\nu} \phi)\,,
\]
for arbitrary solutions $\psi$, $\phi$ of two-component KG equation. The covariant position operator can be defined
as follows:
\begin{eqnarray}\label{x-feynmann-bunge}
\hat x_D^{\mu}=x^{\mu}+\frac{\Sigma^{\mu\alpha}\hat p_\alpha}{2\hat p^2}+\frac{i\hbar(\gamma^5-1)\hat p^\mu}{2\hat
p^2}=\\
x^{\mu}+\frac{i\hbar\gamma^{\mu}}{2\hat p^2}(\gamma\hat p)+\frac{i\hbar\gamma^5\hat p^\mu}{2\hat p^2} \,,\nonumber
\end{eqnarray}
where $\gamma_5=-i\gamma^0\gamma^1\gamma^2\gamma^3$. Again, one can check that matrix elements in two theories coincide
\[
\int d^3x \Psi[\psi]^\dagger\hat{x}_D^{\mu}\Phi[\phi]=(\psi, \hat{x}_{rp}^{\mu} \phi)\,.
\]
As a result, the manifestly-covariant operators $\hat j_D^{\mu\nu}$ and $\hat x_D^{\mu}$ of the Dirac equation represent
position ${\bf x}$ and spin ${\bf S}$ (\ref{spin}) of the Frenlel electron (\ref{s.15}).  Their mean values can be computed as follows
\begin{eqnarray}\label{final}
\langle \Psi,\hat X^i\Phi \rangle=\frac12\mbox{Re}(\Psi[\psi], [\hat x_{D}^i+\hat x_{D}^{i\dagger}]\Phi[\phi] ),
\\
\langle \Psi,\hat S^i\Phi \rangle=\frac14\epsilon^{ijk}(\Psi[\psi], \hat j_{D}^{jk}\Phi[\phi] )\,.
\end{eqnarray}

\section{Conclusions}

The content and the main results of this work have been described in Introduction. So, here we finish with some
complementary comments.

There are a lot of candidates for spin and position operators of the relativistic electron. Different position
observables coincide when we consider standard quasi-classical limit. So, in absence of a systematically constructed
classical model of an electron it is difficult to understand the difference between these operators. Our approach
allows us to do this, after realizing them at the classical level. As we have seen, various non-covariant, covariant and
manifestly-covariant operators acquire clear meaning in the Lagrangian model of Frenkel electron developed in this work.

Starting with variational formulation we described the relativistic Frenkel electron with the aid of singular
Lagrangian. Equations of motion for the classical model are consistent \cite{DPM2, DPW2} with experimentally tested BMT
equations. We showed that the classical variables of position are non-commutative quantities. Selecting physical-time
parametrization in our model in the case of free electron, we have done canonical quantization procedure. As it should
be, we arrived at quantum mechanics which can be identified with positive-energy part of Dirac theory in the
Foldy-Wouthuysen representation. The Foldy-Wouthuysen mean-position and spin operators correspond to canonical
variables $\tilde x_j$ and $\tilde{s}_j$ of the model, whereas the classical position ${\bf x}$ and spin ${\bf S}$ are
represented by Pryce (d)-operators. Since all variables obey the same equations in the free theory, the question of
which of them are the true position and spin is a matter of convention. The situation changes in interacting theory,
where namely ${\bf x}$ and ${\bf S}$ obey the expected F-BMT equations and thus represent the position and spin.

Concerning the position, in his pioneer work \cite{pryce1948mass}, Pryce noticed that ``except the particles of spin
$0$, it does not seem to be possible to find a definition which is relativistically covariant and at the same time
yields commuting coordinates''. Now we know, why this happens. At the classical level, an accurate account of spin (that
is of Frenkel condition) in a Lagrangian theory yields, inevitably, the relativistic corrections to the classical
brackets of position variables.

It seems to be very interesting to study $\hat{X}_{P(d)}^j$ as the ``true'' relativistic position operator in more
details. The first reason is an interesting modification of quantum interaction between the electron and background
electromagnetic fields coming from non-local interactions $\hat p^\mu\rightarrow \hat p^\mu
-\frac{e}{c}A^{\mu}(\hat{X}_j)$, $F^{\mu\nu}(\hat{X}_j)\hat J_{\mu\nu}$. The second reason is due to its natural
non-commutativity, that could be contrasted with a number of theoretical models where non-commutativity introduced by
hands. We return to these issues in the next paper \cite{DPM2}.

We also quantized our model in an arbitrary parame-\\ terization, keeping the manifest Lorentz-invariance. 
The covariant
quantization gives positive-energy sector of two-component Klein-Gordon equation 
(quantum field theory of two-component
KG has been proposed by Feynman and Gell-Mann \cite{feynman1958fermi-interaction}). We have found a covariant conserved
current for the two-component KG equation, which allows us to define an invariant, positive-definite scalar product
with metric $\rho$ in the space of two-component spinors. The resulting relativistic quantum mechanics represents
one-particle sector of the Feynman-Gell-Mann quantum field theory. Classical spin-plane invariant variables $p^\mu$,
$S^\mu$ and $J^{\mu\nu}$ produce manifestly-covariant operators.

The square root of metric, $V=\rho^{1/2}$, defines the map from canonical to covariant formulations. This allows us to
establish relativistic covariance of canonical formalism: scalar product and mean values of operators of canonical
formalism can be computed using the corresponding quantities of covariant formalism, see the proposition of subsection
\ref{covinv}. And back, the transformation $V$ allows us to interpret the results of covariant quantization in terms of
one-particle observables of an electron in the FW representation (see Table
\ref{tabular:non-covariantFW-covariant-quantization}). The relativistic-position operator $\hat x^\mu_{rp}$ is
non-Hermitian and does not correspond to a physical observable. However, pseudo-Hermitian part of $\hat x^j_{rp}$
coincides with image of physical-position operator $\hat{x}^j_V=V^{-1}\hat{X}^iV$.

Our classical model may provide a unification in modern issues of quantum observables in various theoretical and
experimental setups [31-46]. Since the model constructed admits an interaction with electro-magnetic and gravitational
fields, one can try to extend the obtained results beyond the free relativistic electron.

\section{Acknowledgments}
This work has been supported by the Brazilian foundation CNPq. AMPM thanks CAPES for the financial support ( Programm
PNPD/2011).

\section*{Appendix 1. Some identities}\label{sec:app1}
\begin{eqnarray}\label{aa1}
\eta^{\mu\nu}=(-, +, +, +), \quad \epsilon^{0123}=1, \quad \epsilon_{0123}=-1, \cr
\epsilon^{abcd}\epsilon_{ab\mu\nu}=-2(\delta^c{}_\mu\delta^d{}_\nu-\delta^c{}_\nu\delta^d{}_\mu). \nonumber\\
\label{aa2}
\epsilon^{\mu abc}\epsilon_{\mu ijk}=-[\delta^a{}_i(\delta^b{}_j\delta^c{}_k-\delta^b{}_k\delta^c{}_j)+
{\rm cycle}(ijk)]. \nonumber
\end{eqnarray}

Given quantities $J^{\mu\nu}=-J^{\nu\mu}$ and $p^\mu$, we define the vectors
\begin{eqnarray}\label{aa3}
s^\mu=\frac{1}{4\sqrt{-p^2}}\epsilon^{\mu\nu\alpha\beta}p_\nu J_{\alpha\beta}, \quad \mbox{then} \quad s^\mu p_\mu=0,
\end{eqnarray}
\begin{eqnarray}\label{aa4}
\Phi^\mu=J^{\mu\nu}p_\nu, \qquad \mbox{then} \qquad \Phi^\mu p_\mu=0. \nonumber
\end{eqnarray}
Then both $J^{\mu\nu}$ and its dual, ${}^*J^{\mu\nu}=\frac{1}{2} \epsilon^{\mu\nu ab}J_{ab}$, can be decomposed on
these vectors
\begin{eqnarray}\label{aa5}
J^{\mu\nu}=\frac{\Phi^\mu p^\nu-\Phi^\nu p^\mu}{p^2}-\frac{2}{\sqrt{-p^2}}\epsilon^{\mu\nu ab}p_as_b, \nonumber
\end{eqnarray}
\begin{eqnarray}\label{aa6}
\epsilon^{\mu\nu ab}J_{ab}=4\frac{p^\mu s^\nu-p^\nu s^\mu}{\sqrt{-p^2}}-\frac{2}{p^2}\epsilon^{\mu\nu ab}p_a\Phi_b. \nonumber
\end{eqnarray}
we have the identity
\begin{eqnarray}\label{aa7}
s^\mu s_\mu=-\frac{1}{4p^2}(J^{\mu\nu}p_\nu)^2+\frac{1}{8}J^{\mu\nu}J_{\mu\nu}. \nonumber
\end{eqnarray}
If, in addition to this, $J^{\mu\nu}$ obeys
\begin{eqnarray}\label{aa8}
J^{\mu\nu}p_\nu=0, \nonumber
\end{eqnarray}
then $J^{\mu\nu}$ and $S^\mu$ turn out to be equivalent
\begin{eqnarray}\label{aa9}
s^\mu=\frac{1}{4\sqrt{-p^2}}\epsilon^{\mu\nu\alpha\beta}p_\nu J_{\alpha\beta}, \qquad
J^{\mu\nu}=-\frac{2}{\sqrt{-p^2}}\epsilon^{\mu\nu\alpha\beta}p_\alpha s_\beta, \nonumber
\end{eqnarray}
and obey the identities
\begin{eqnarray}\label{aa10}
\epsilon^{\mu\nu ab}J_{ab}=4\frac{p^\mu s^\nu-p^\nu s^\mu}{\sqrt{-p^2}}, \nonumber
\end{eqnarray}
\begin{eqnarray}\label{aa11}
s^\mu s_\mu=\frac{1}{8}J^{\mu\nu}J_{\mu\nu}. \nonumber
\end{eqnarray}
In the rest system of $p^\mu$, $p^\mu=(p^0, \vec 0)$, $\sqrt{-p^2}=|p^0|=mc$ we have
\begin{eqnarray}\label{aa12}
s^0=0, \qquad s^i=\frac{p^0}{4|p^0|}\epsilon^{ijk}J_{jk}. \nonumber
\end{eqnarray}
The last equality explains our normalization for the BMT vector $s^\mu$ , Eq. (\ref{aa3}).

\section*{Appendix 2. General solution to equations of motion \label{subsec:min-action-lagrange-eqs}}
{\bf Lagrangian equations.} Variation of the minimal action (\ref{s.15}) implies the equations
\begin{eqnarray}\label{la.1}
\frac{\delta S}{\delta g_4}=0 ~: \quad \omega^2=a_4, \quad \Rightarrow \quad (\omega\dot\omega)=0,
\end{eqnarray}
\begin{eqnarray}\label{la.2}
\frac{\delta S}{\delta x} &=& 0 ~: ~ -mc\left(\frac{N\dot x^\mu}{\sqrt{-\dot xN\dot x}}\right)\dot{}=0, \quad \Rightarrow
\cr
 &-& mc\frac{N\dot x^\mu}{\sqrt{-\dot xN\dot x}}=p^\mu=\mbox{const}, \quad \Rightarrow \quad 
\cr 
(p\omega) &=& 0, \quad
p^2=-(mc)^2, \quad \Rightarrow \quad (p\dot\omega)=0, \qquad \qquad \qquad
\end{eqnarray}
\begin{eqnarray}\label{la.3}
&\,&\frac{\delta S}{\delta\omega}=0: 
\cr &\,&\sqrt{a_3}\left(\frac{N\dot\omega^\mu}{\sqrt{\dot\omega
N\dot\omega}}\right)\dot{}+\frac{\sqrt{a_3}(\omega\dot\omega)}{\omega^2\sqrt{\dot\omega N\dot\omega}}N\dot\omega^\mu
+\cr 
&\,&\frac{(\omega\dot x)}{\omega^2}\frac{mcN\dot x^\mu}{\sqrt{-\dot xN\dot x}}+g_4\omega^\mu=0.
\end{eqnarray}
Using the consequences pointed in Eqs. (\ref{la.1}) and (\ref{la.2}), we simplify the equation (\ref{la.3})
\begin{eqnarray}\label{la.4}
\sqrt{a_3}\left(\frac{\dot\omega^\mu}{\sqrt{\dot\omega^2}}\right)\dot{}-\frac{(\omega\dot
x)}{a_4}p^\mu+g_4\omega^\mu=0. \nonumber
\end{eqnarray}
Contraction of this equation with $\omega^\mu$ gives the expression for $g_4$
\begin{eqnarray}\label{la.5}
g_4=\frac{\sqrt{a_3}\sqrt{\dot\omega^2}}{a_4}, \nonumber
\end{eqnarray}
whereas contraction with $p^\mu$ implies $(\omega\dot x)=0$. Collecting all this, the initial Lagrangian equations can
be presented in the equivalent form
\begin{eqnarray}\label{la.6}
\left(\frac{\dot x^\mu}{\sqrt{-\dot x^2}}\right)\dot{}=0,
\end{eqnarray}
\begin{eqnarray}\label{la.7}
\left(\frac{\dot\omega^\mu}{\sqrt{\dot\omega^2}}\right)\dot{}+\frac{\sqrt{\dot\omega^2}}{a_4}\omega^\mu=0,
\end{eqnarray}
\begin{eqnarray}\label{la.8}
\omega^2=a_4, \quad (\omega\dot x)=0.
\end{eqnarray}
We have second-order equations (\ref{la.6}) and (\ref{la.7}). Besides, there are presented two Lagrangian constraints
(\ref{la.8}).

{\bf General solution} to equations (\ref{la.6})-(\ref{la.8}) reads
\begin{eqnarray}\label{lo.8}
x^\mu=x^\mu_0+p^\mu\lambda_1(\tau),
\end{eqnarray}
\begin{eqnarray}\label{lo.9}
\omega^\mu=\sqrt{\frac{a_4}{a_3}}A^\mu\sin f(\tau)+\sqrt{\frac{a_4}{a_3}}B^\mu\cos f(\tau),
\end{eqnarray}
where $\lambda_1$ and $f$ are arbitrary functions of the evolution parameter. Constants of integration obey the
restrictions
\begin{eqnarray}\label{lo.10}
p^2 &=& -(mc)^2, \quad (pA)=(pB)=0, \quad A^2=a_3,\cr 
B^2 &=& a_3, \quad (AB)=0.
\end{eqnarray}

Eq. (\ref{lo.8}) determines straight line (as geometric place of points) in Minkowski space, whereas (\ref{lo.9}) is an
ellipse which lies on the plane of inner space formed by the vectors $A^\mu$ and $B^\mu$. Due to the arbitrary
functions $\lambda(\tau)$ and $f(\tau)$, evolution along the trajectories is not specified, as it should be in a
reparametrization invariant theory.

{\bf General solution to Hamiltonian equations.} Hamiltonian formulation leads to the same result. Hamiltonian
constraints and equations written in section \ref{sec:min-action-hamilton} do not determine the multipliers $\lambda_1$
and $\lambda_4$. As a consequence, the variable $g_4(\tau)$ can not be determined neither with the constraints nor with
the dynamical equations. This implies the functional ambiguity in solutions to the equations of motion for the basic
variables $x^\mu$, $\omega^\mu$ and $\pi^\mu$: besides the integration constants, solution depends on these arbitrary
functions.

Denoting
\begin{eqnarray}\label{ha.11}
f(\tau)=\sqrt{\frac{a_4}{a_3}}\int d\tau g_4, \nonumber
\end{eqnarray}
general solution to the Hamiltonian equations is given by Eqs. (\ref{lo.8})-(\ref{lo.10}) and
\begin{eqnarray}\label{ha.13}
\pi^\mu=A^\mu\cos f(\tau)-B^\mu\sin f(\tau). \nonumber
\end{eqnarray}

{\bf Covariant dynamics in proper-time\\
parametrization.} 

The physical variables obey to non degenerate equations, but
they are not manifestly covariant. The standard way to work with non-degenerated equations keeping covariance is to fix
parametrization to be proper time of the particle, $x^\mu=x^\mu(s)$. Here $s$ is the time measured in the
instantaneous rest frame. As the proper time coincides with interval between the particle positions,
in the proper-time parametrization we have the relation\footnote{In an arbitrary parametrization we have $(\dot
x^\mu(\tau))^2=-c^2\dot s^2(\tau)$, that is nothing interesting.}
\begin{eqnarray}\label{ha.21}
(\dot x^\mu(s))^2=-c^2. \nonumber
\end{eqnarray}
This equation together with Eq. (\ref{ha.9}) fixes $\lambda_1=\frac{1}{m}$, so we arrive at the deterministic equations
\begin{eqnarray}\label{ha.22}
\dot x^\mu(s)=\frac{p^\mu}{m}, \quad \dot p^\mu(s)=0; \quad  p^2=-(mc)^2, \nonumber
\end{eqnarray}
with the solution being
\begin{eqnarray}\label{ha.23}
x^\mu=x^\mu_0+\frac{p^\mu}{m}s, \quad p^2=-(mc)^2, \quad p^\mu=\mbox{const}. \nonumber
\end{eqnarray}
As before, physical dynamical variables $x^i(t)$ obtained from $x^\mu(s)$ excluding the parameter $s$.

Spin-sector is described either by Eq. (\ref{pha.12}) or  by Eq. (\ref{pha.14}).

\end{document}